\documentclass[12pt]{iopart}
\usepackage{graphicx}
\usepackage{subfigure}
\usepackage{cite}
\usepackage{bm}

\begin{document}

\title[Application of RLLM in CDCC]{Application of the Lagrange-mesh method in continuum-discretized coupled-channel calculations}

\author{Wendi Chen$^1$, Hairui Guo$^2$, Tao Ye$^2$, Yangjun Ying$^2$ Weili Sun$^2$ and Yinlu Han$^3$}

\address{$^1$ School of Physics, Beihang University, Beijing 100191, China}
\address{$^2$ Institute of Applied Physics and Computational Mathematics, Beijing 100094, China}
\address{$^3$ Key Laboratory of Nuclear Data, China Institute of Atomic Energy, Beijing 102413, China}
\ead{guo$\_$hairui@iapcm.ac.cn}
\vspace{10pt}
\begin{indented}
\item[] 25 April 2022
\end{indented}

\begin{abstract}
We apply the Lagrange-mesh method to discretize continuum states of weakly bound nuclei for continuum-discretized coupled-channel (CDCC) calculations of three-body breakup reactions. This discretization method is compared with the bin method, which is regarded as the standard continuum discretization method, for the $d$ and $^6$Li induced reactions. An improved Numerov algorithm is used to solve the coupled channels equations, which permits a fast integration of equations and a convenient treatment of the closed channels. A new CDCC model code named \emph{CDCC-R} is developed. In all cases, the combination of the Lagrange-mesh method and the Numerov algorithm shows high efficiency and accuracy for the CDCC calculations of the elastic scattering and breakup reactions. Especially, various numerical and physical aspects are discussed for $^6$Li induced reactions. The coupling effects of the continuum states with high orbital angular momentum and closed channels are discussed. Moderate effects are found in the calculations for elastic scattering and breakup reaction when $F$- and $G$-wave continuum states are included in CDCC model space for $^6$Li induced reactions at the incident energies well above the Coulomb barrier. The closed channel effect is found to be indispensable for $^6$Li breakup reaction calculation when the incident energy is around the Coulomb barrier.
\end{abstract}

%
% Uncomment for keywords
%\vspace{2pc}
%\noindent{\it Keywords}:
%
% Uncomment for Submitted to journal title message
%\submitto{\JPG}
%
% Uncomment if a separate title page is required
%\maketitle
%
% For two-column output uncomment the next line and choose [10pt] rather than [12pt] in the \documentclass declaration
%\ioptwocol
%

\section{\label{s-intro}Introduction}

The continuum-discretized coupled-channel (CDCC) method is a powerful tool to investigate the reaction mechanisms of weakly bound nuclei induced reactions, which are influenced significantly by the breakup effect \cite{Yahiro2012}. Although it is originally designed to describe deuteron-nucleus reactions \cite{Yahiro1986,Austern1987,ChauHuu-Tai2006}, CDCC has succeeded in reproducing and analysing the scattering data of not only stable nuclei but also unstable nuclei, such as $^{6,7}$Li\cite{Sakuragi1986,Matsumoto2003,Diaz-Torres2003a,Beck2007,Camacho2019}, $^{7,11}$Be\cite{Keeley2002,Diaz-Torres2002,Keeley2010}, $^{8}$B\cite{Tostevin2001,Lubian2010}, $^{17}$F\cite{Mazzocco2010,Kucuk2012}, etc. Nowadays CDCC has been extensively applied in the studies of elastic scattering \cite{Yahiro1986,ChauHuu-Tai2006}, polarization potential \cite{Mackintosh2004}, breakup reaction \cite{Tostevin2001}, transfer reaction \cite{Shrivastava2006} and fusion reaction\cite{Diaz-Torres2002,Diaz-Torres2003a,Beck2007,Camacho2019} for the weakly bound nuclei induced reactions. Some researches based on the CDCC framework for a further description of the reactions have been published, such as microscopic CDCC \cite{Sakuragi1986,Descouvemont2013}, four body CDCC \cite{Kamimura2005,Watanabe2012}, extended CDCC including the effect of core excitation \cite{Summers2006,Summers2007} or target excitation \cite{Gomez-Ramos2017}.

CDCC solves the three-body problem by projecting the full wave function onto the model space expanded by a set of internal wave functions of the weakly bound nucleus. Therefore it is an important issue in CDCC that the continuum states of the weakly bound nucleus should be well represented by a finite number of internal states. There are two kinds of discretization methods so far. One is the bin method \cite{Kawai1986,Sakuragi1986,Austern1987,Piyadasa1999}, in which the continuum is truncated and divided into a finite number of intervals in energy or momentum space. The average of exact scattering wave functions in the interval (average method) or the scattering wave function at the centre of the interval (midpoint method) is used to represent the continuum in that interval. It has been proven that the average and midpoint methods yield the same converging $S$-matrix elements when the width of the interval is small enough \cite{Piyadasa1999}. The bin method is generally regarded as the standard continuum discretization method and its result can be used as the benchmark for CDCC calculation. Another discretization method is the pseudo-state (PS) method, which provides the discrete state wave functions via diagonalizing the internal Hamiltonian of the weakly bound nucleus with a set of square-integrable basis functions. Gaussian basis \cite{Hiyama2003a} and transformed harmonic oscillator basis \cite{Perez-Bernal2001,Moro2002,Moro2009} are both alternatives.

However, it requires a large amount of calculation in CDCC with the above methods because of the numerical integration for coupling matrix elements. In the present work, the Lagrange-mesh method is applied to discretize the continuum states of weakly bound nuclei and calculate their bound states. With this method, the CDCC coupling matrix elements can be calculated by Gauss-quadrature approximation, which reduces the amount of calculation remarkably. The validity of applying Lagrange-mesh method in CDCC has been introduced by T. Druet et al. \cite{Druet2010} for $d$($p$+$n$)+$^{58}$Ni reaction and by T. Druet and P. Descouvemont \cite{Druet2012} for $^{11}$Be($^{10}$Be+$n$)+$^{64}$Zn reaction. However, in their studies, the CDCC equations were both solved by $R$-matrix approach. The $R$-matrix approach is a universal and accurate solution for scattering problems \cite{Descouvemont2016,Shubhchintak2019,Lei2020}, but it is very time-consuming, especially when the number of coupled channels is large. In the CDCC calculations for heavy weakly bound nuclei induced reactions, the number of channels is typically larger than 100 to well include the breakup effect, resulting in a large time consumption with the $R$-matrix approach. Therefore, an improved Numerov algorithm \cite{Yang1980} is adopted by us to solve the coupled channel equations, which permits a fast integration of equations and a convenient treatment of the closed channels. Accordingly, we develop a new CDCC model code named \emph{CDCC-R}.

In this paper, the following reactions are chosen to check the validity of our calculations:

(1) $d$+$^{58}$Ni at EL=80.0 MeV.

(2) $^{6}$Li+$^{12}$C at EL=168.6 and 178.0 MeV.

(3) $^{6}$Li+$^{59}$Co at EL=12.0, 17.4 and 18.0 MeV.

EL represents the incident energy of the projectile in the laboratory system. The CDCC results calculated with the bin method are regarded as the benchmark for comparison. Various and detailed calculations will be made for $^6$Li induced reactions. The coupling effects of the continuum states with high orbital angular momentum and closed channels will be discussed. It will be also shown that the combination of the Lagrange-mesh method and the Numerov algorithm can provide a suitable and efficient representation for the continuum states of a two-body system and a quick calculation for CDCC equations.

The paper is organized as follows. We briefly present the formalisms of CDCC, bin method, Lagrange-mesh method, elastic scattering and breakup calculation in Sec. \ref{sec-form}. A benchmark calculation of the reaction $d$+$^{58}$Ni at EL=80.0 MeV is given in Sec. \ref{sec-benchmark}. The applications of the Lagrange-mesh method to $^6$Li induced reactions and related discussions are presented in Sec. \ref{sec-appl}. The summary and conclusion are given in Sec. \ref{sec-conc} eventually.

\section{Theoretical formalisms}\label{sec-form}

\subsection{Outline of CDCC}\label{sec-form-1}

There are many references which have detailed descriptions of the CDCC theory \cite{Kawai1986,Sakuragi1986,Austern1987}. Here we only present a brief introduction and define some notations. The weakly bound nucleus is always the projectile in this paper. $j-j$ coupling scheme is adopted.

The weakly bound projectile is treated as two-body system, consisting of a core particle ($c$) and a valence particle ($v$). The two-body Hamiltonian of the projectile is written as
\begin{equation}\label{e-Ham-proj}
  H_P=T_{c-v}+V_{c-v},
\end{equation}
where $T_{c-v}$ and $V_{c-v}$ represent the relative motion kinetic energy and interaction of $c$-$v$ system respectively. In CDCC formalism, the continuum states of the projectile are represented by a finite number of discretized states. For the bound states and discretized states, the projectile wave functions are expressed on the same footing as
\begin{equation}\label{e-wf-proj}
\eqalign{
  \phi _{l,j,I}^{n} (\vec{r})&=\frac{\varphi _{l,j,I}^{n}\left( r \right)}{r}\left[ \left[ i^lY_l\left( \bm{\Omega_{r}} \right) \otimes \chi_{s _v} \right] _j \otimes \chi_{s _c} \right] _I,1\le n\le N_{l,j,I}, \\
  H_P \phi _{l,j,I}^{n} &=\varepsilon _{l,j,I}^n \phi _{l,j,I}^{n},
}
\end{equation}
where the square bracket represents the angular momentum coupling. $\vec{r}$ is the coordinate of the valence particle relative to the core particle. $\bm{\Omega_{r}}$ is the solid angle of $\vec{r}$. $l$ is the orbital angular momentum for the $c$-$v$ relative motion. $\chi_{s _v}$ and $\chi _{s_c}$ represent the spinors of $v$ and $c$ respectively. The spins of $v$ and $c$ are $s_v$ and $s_c$ respectively. $I$ is the angular momentum of projectile. $\varphi _{l,j,I}^{n}(r)$ is the radial wave function for $c$-$v$ system and it must be square-integrable for CDCC calculation. $N_{l,j,I}$ denotes the number of states within the angular momentum coupling scheme $(l,j,I)$. $\varepsilon _{l,j,I}^n$ is the energy of $\phi _{l,j,I}^{n}$.

In the present work, we ignore the spin and excitation of the target nucleus, then the total Hamiltonian of the system can be expressed as
\begin{equation}\label{e-Ham-sys}
\eqalign{
  H&=H_P+T_{P-T}+V_{c-T} (\vec{R}+f_c \vec{r})+V_{v-T} (\vec{R}+f_v \vec{r}), \\
  f_c&=\frac{m_v}{m_c+m_v}, f_v=-\frac{m_c}{m_c+m_v},
}
\end{equation}
where $T_{P-T}$ denotes the kinetic energy of projectile-target relative motion. $V_{c-T}$ and $V_{v-T}$ are the interactions of $c$-target and $v$-target systems respectively. $\vec{R}$ is the relative coordinate between projectile and target. $m_c$ and $m_v$ are the masses of core and valence particles respectively. In CDCC formalism,  the total wave function with total angular momentum $J$ and parity $\pi$ is expanded over the projectile internal wave functions as
\begin{eqnarray}\label{e-wf-sys}
    \Phi ^{J,\pi} &=\sum_{\beta}{\frac{u_{\beta}^{J,\pi}\left( R \right)}{R}\psi _{\beta}^J}, \\
    \psi _{\beta} ^J &=\left[ i^LY_L ( \bm{\Omega_{R}}  ) \otimes \phi _{l,j,I}^{n} \right] _J,
\end{eqnarray}
where $\beta$ denotes all the quantum numbers necessary to define the channel $\beta =\{ n,l,j,I,L \}$. $\bm{\Omega_{R}}$ is the solid angle of $\vec{R}$. $L$ is the orbital angular momentum for the projectile-target relative motion. $u_{\beta}^{J,\pi}$ represents the projectile-target radial relative motion in $\beta$ channel with $J$ and $\pi$.

By projecting the Schr\"{o}dinger equation $H \Phi = E \Phi $ onto $\psi _{\beta} ^J$,  a set of coupled channel equations are generated to determine $u_{\beta}^{J,\pi}$, that is
\begin{equation}\label{e-CDCC}
\eqalign{
  \left[ \frac{\hbar ^2}{2\mu _{P-T}}\left( -\frac{d^2}{dR^2}+\frac{L\left( L+1 \right)}{R^2} \right) +V_{\beta ,\beta}^{J,\pi} (R)+\varepsilon _{\beta}-E \right] u_{\beta}^{J,\pi}(R) \\
  =-\sum_{\beta '\ne \beta}{V_{\beta ,\beta '}^{J,\pi} (R)}u_{\beta '}^{J,\pi}(R),
}
\end{equation}
where
\begin{equation}\label{e-CDCC-Vcc}
  V_{\beta ,\beta '}^{J,\pi} (R)=\left< \psi _{\beta}^{J,\pi} \right|V_{c-T} + V_{v-T} \left| \psi _{\beta '}^{J,\pi} \right> .
\end{equation}
$\varepsilon _{\beta}$ is the projectile internal energy in the $\beta$ channel. $\mu_{P-T}$ is the projectile-target reduced mass. If $V_{c-T}$ and $V_{v-T}$ are only central potentials and rewritten by multipole expansion as
\begin{equation}\label{e-V-expand}
  V_{c-T}+V_{v-T}=\sum_{\lambda}{ (2 \lambda +1) V_{\lambda}( R,r ) P_{\lambda}( \cos \theta )},\cos \theta=\bm{\Omega_{R}} \cdot \bm{\Omega_{r}},
\end{equation}
one can obtain the coupling matrix elements as
\begin{equation}\label{e-Vcc}
\eqalign{
 V_{\beta ,\beta '}^{J,\pi} (R)=& i^{L'+l'-L-l}(-1)^{s_c+s_v+l'+l+j+j'+I-I'-J}\hat{l}\hat{l}'\hat{j}\hat{j}'\hat{I}\hat{I}'\hat{L}\hat{L}'
\\
&\times \sum_{\lambda}{\left( 2\lambda +1 \right)}
\left\{ \begin{array}{c}
	l' \quad        j' \quad		s_v\\
	j  \quad        l  \quad	    \lambda\\
\end{array} \right\}
\left\{ \begin{array}{c}
	j' \quad		I' \quad	    s_c\\
	I  \quad		j  \quad		\lambda\\
\end{array} \right\}
\left\{ \begin{array}{c}
	I' \quad		L' \quad		J\\
	L  \quad		I  \quad		\lambda\\
\end{array} \right\}
\\
&\times \left( \begin{array}{c}
	L' \quad		\lambda   \quad		L\\
	0  \quad		0         \quad		0\\
\end{array} \right)
\left( \begin{array}{c}
	l' \quad		\lambda   \quad		l\\
	0  \quad		0         \quad		0\\
\end{array} \right)
\int_0^{+\infty}{\varphi _{l,j,I}^{n}V_{\lambda}\left( R,r \right) \varphi _{l',j',I'}^{n'}dr}
}
\end{equation}
where $\hat{x}$=$\sqrt{2x+1}$. $3j$ and $6j$ symbols appear as usual. This expression of $V_{\beta, \beta'}^{J,\pi}$ is equivalent to those given in Refs. \cite{Nishioka1984,Thompson1988}, although the comparison requires some angular momentum algebra. It can be derived from the $3j$ symbols in Eq. (\ref{e-Vcc}) that the maximum order of multipole expansion for $V_{c-T}$ and $V_{v-T}$, $\lambda _{\max}$, should not be larger than 2$l_{\max}$, where $l_{\max}$ is the maximum $l$ of projectile continuum included in CDCC calculation.

\subsection{The bin method}\label{sec-form-2}

In the present paper, the calculated results with the bin method \cite{Kawai1986,Sakuragi1986} are regarded as the benchmark for comparison. The wave number of $c-v$ system is defined as
\begin{equation}\label{e-kcv}
  k=\frac{\sqrt{2\mu _{c-v}\varepsilon}}{\hbar},
\end{equation}
where $\mu _{c-v}$ and $\varepsilon$ are the reduced mass and the relative energy in centre of mass system of $c-v$ system respectively. In the bin method, the continuum is truncated by restricting $l$ and wave number $k$ as
\begin{equation}\label{e-bin-trun}
  l \leq l_{\max} \quad \mathrm{and} \quad k \leq k_{\max}.
\end{equation}
The discretized state wave functions $\phi _{l,j,I}^n$ are generated by dividing the continuum states $\{ \phi _{l,j,I}^{c}(k,\vec{r}); 0<k<k_{\max} \}$ into finite bins $\{ [ k_{i-1}, k_i ] ; 1\leq i \leq N_{l,j,I}^c \} $ and then averaging continuum state wave function $\phi _{l,j,I}^{c}$ in each bin. That is
\begin{eqnarray}\label{e-wf-bin}
  &\phi _{l,j,I}^{c} (k,\vec{r}) =\frac{\varphi _{l,j,I}^{c}(k,r)}{r}\left[ \left[ i^lY_l\left( \Omega _r \right) \otimes \chi _{s_v} \right] _j\otimes \chi _{s_c} \right] _I, \\
  &\varphi _{l,j,I}^{c}\left( k,r\right) \rightarrow \sqrt{\frac{2}{\pi}}\left[ \cos \delta _{l}^{j,I}F_l\left( kr \right) +\sin \delta _{l}^{j,I}G_l\left( kr \right) \right], \label{e-continuum-rwf}\\
  &\phi _{l,j,I}^{n} =\frac{1}{\sqrt{N_w}}\int\limits_{k_{i-1}}^{k_i}{w\left( k \right) \phi _{l,j,I}^{c}\left( k,\vec{r} \right) dk}, \label{e-wf-bin-form}\\
  &N_w=\int\limits_{k_{i-1}}^{k_i}{\left| w\left( k \right) \right|^2dk},n =i+N_{l,j,I}^b,
\end{eqnarray}
where $N_{l,j,I}^b$ is the number of bound states in angular momentum coupling scheme $(l,j,I)$. $N_{l,j,I}^b$ + $N_{l,j,I}^c$ = $N_{l,j,I}$. $F_l(kr)$ and $G_l(kr)$ are regular and irregular Coulomb wave functions respectively \cite{Thompson2010}. $\delta _{l}^{j,I}$ is the scattering phase shift of $c$-$v$ system. The weight function $w$=1 is used for non-resonance states and $w$=$\sin \delta _{l}^{j,I}$ is adopted for resonance states \cite{Sakuragi1986,Thompson1988}. In practical calculation, $\phi _{l,j,I}^{n}$ is truncated at a sufficiently large radius ($r_{bin}$) to obtain converging coupling matrix elements and solutions of the coupled channel equations.

\subsection{Brief introduction of the Lagrange-mesh method}\label{sec-form-3}

The Lagrange-mesh method is an approximate variational method and has been applied in many different physical fields \cite{Baye2015}. Its basis functions and relative integration are associated with Gauss quadrature approximation and therefore a satisfying high accuracy can be obtained with a small amount of computation.

For the half-infinite interval $\left[ 0,+\infty \right] $, the regularized Lagrange-Laguerre mesh method (RLLM) is adopted and its $N$ basis functions are defined as
\begin{equation}\label{e-RLLMbf}
f_i\left( r \right) =\frac{\left( -1 \right) ^i}{\sqrt{h x_i}}\frac{L_N\left( r/h \right)}{r-hx_i}re^{-r/2h},
\end{equation}
where $L_N$ is the Laguerre polynomial of degree $N$. $x_i$ corresponds to the zero of $L_N$, that is,
\begin{equation}\label{e-zeros}
  L_N(x_i)=0, i=1,2,...,N.
\end{equation}
$h$ is a scaling parameter, adopted to the typical size of the system. These functions are used to diagonalize the Hamiltonian with Gauss quadrature\cite{Baye2015} and generate a set of eigenfunctions $\varphi _i$.
\begin{equation}\label{e-RLLMeigen}
  \varphi _i=\sum_{j=1}^N{c_{j}^{i}f_j}.
\end{equation}

The $\varphi _i$ with negative eigenvalue ($\varepsilon _i$ <0) are the bound state radial wave functions and those with positive eigenvalue ($\varepsilon _i$ >0) are regarded as pseudo-state radial wave functions. Similarly to the bin method, the truncation is made by restricting $l$ and $\varepsilon _i$ as
\begin{equation}\label{e-RLLM-trun}
  l \leq l_{\max} \quad \mathrm{and} \quad \varepsilon _i \leq \varepsilon _{\max}=\frac{\hbar ^2k_{\max}^{2}}{2\mu _{c-v}}.
\end{equation}

The Lagrange condition reads
\begin{equation}\label{e-RLLMcondition}
f_i\left( hx_j \right) =\frac{1}{\sqrt{h\lambda _j}}\delta _{i,j},
\end{equation}
where $\lambda _j$ is the Gauss quadrature weight corresponding to $x_j$. The overlap and integration with local potential $V(r)$ can be obtained with Gauss quadrature efficiently as
\begin{equation}\label{e-RLLMoverlap}
\eqalign{
\left< f_i \mid f_j \right> & \approx \delta _{i,j},
\\
\left< f_i \right|V\left( r \right) \left| f_j \right> & \approx V\left( hx_i \right) \delta _{i,j}.
}
\end{equation}
Therefore, the integration term in Eq. (\ref{e-Vcc}) can be calculated with a few potential values at the mesh points as
\begin{equation}\label{e-RLLMinte}
\int{\varphi _iV \varphi _jdr}=\sum_{k=1}^N{c_{k}^{i}c_{k}^{j}V ( hx_k )}.
\end{equation}
The accuracy of the Gauss approximation in the Lagrange-mesh method has been discussed in many references \cite{Baye2015,Baye2011,Baye2002}. Appropriate RLLM parameters will be given in Sec. \ref{sec-appl} to achieve convergence in CDCC calculations.

\subsection{Elastic scattering and breakup calculation}\label{sec-form-4}

By solving Eq. (\ref{e-CDCC}), one can obtain the $S$-matrix $S^{J,\pi}_{\beta,\beta'}$. For the incident particle with spin $I_0$, the elastic scattering angular distribution is calculated as
\begin{equation}\label{e-elastic}
\frac{d\sigma _{\mathrm{el}}}{d\Omega _R}=\frac{1}{2I_0+1}\sum_{m_Im_I'M_LM_L'}{\left| f_C\left( \theta \right) \delta _{m_I,m_I'}\delta _{M_L,M_L'}+f_{m_Im_I'M_LM_L'}\left( \Omega _R \right) \right|^2},
\end{equation}
\begin{equation}\label{e-fN}
\eqalign{
f_{m_Im_I'M_LM_L'}\left( \Omega _R \right)= &i\frac{\sqrt{\pi}}{K_0}\sum_{L_0L_0'J\pi}{\begin{array}{c}
	\hat{L}_0\\
\end{array}}\exp \left\{ i\left( \sigma _{L_0}^{C}+\sigma _{L_0'}^{C} \right) \right\} \left( \delta _{L_0,L_0'}-S_{\beta _0,\beta _0'}^{J,\pi} \right)
\\
&\times \left< I_{0}m_{I}L_{0}0\mid Jm_{I} \right> \left< I_0m_I'L_0'M_L'\mid Jm_{I} \right> Y_{L_0'}^{M_L'}\left( \Omega _R \right),
}
\end{equation}
where $\beta _0=\{1,l_0,j_0,I_0,L_0\}$ is an entrance channel and $\beta _0'=\{1,l_0,j_0,I_0,L_0'\}$. $f_C$ is the Coulomb scattering amplitude. $K_0$ is the wave number of entrance channel. $\sigma_L^C$ is the Coulomb scattering phase shift.

Following the method in Ref. \cite{Matsumoto2003}, the scattering $S$-matrix from an entrance channel $\beta _0=\{1,l_0
,j_0,I_0,L_0\}$ to a breakup configuration with quantum numbers $\gamma=\{ l,j,I,L \}$ and wave number $k$ can be approximated as
\begin{equation}\label{e-CDCC-Smatrix}
  S_{\gamma,\beta _0}^{J,\pi}\left( k \right) =\sum_n{\left< \phi _{l,j,I}^{n} \mid \phi _{l,j,I}^{c}\left( k \right) \right> S_{\beta ,\beta _0}^{J,\pi}},
\end{equation}
The differential breakup cross section $d \sigma _{BU} / dk$ for the continuum states with angular momentum coupling scheme $(l,j,I)$ is calculated as
\begin{equation}\label{e-dbudk}
\eqalign{
& \frac{d\sigma _{BU}(l,j,I)}{dk}=\sum_{J,\pi}{\frac{d\sigma _{BU}^{J,\pi}(l,j,I)}{dk}}
\\
& \frac{d\sigma _{BU}^{J,\pi}(l,j,I)}{dk}=\frac{\pi}{K_{0}^{2}}\frac{1}{2I+1}\sum_{\beta _0L}{\left( 2J+1 \right) \left| S_{\gamma ,\beta _0}^{J,\pi}\left( k \right) \right|^2}.
}
\end{equation}
The breakup cross section can be obtained as
\begin{equation}\label{e-cs-bu}
  \sigma _{BU}=\sum_{l,j,I}{\int{\frac{d\sigma _{BU}(l,j,I)}{dk}dk}}.
\end{equation}

The overlap in Eq. (\ref{e-CDCC-Smatrix}) can be obtained analytically in bin method with the definition in Eq. (\ref{e-wf-bin-form}) as
\begin{equation}\label{e-Bin-overlap}
\eqalign{
\left< \phi _{l,j,I}^{n} \mid \phi _{l,j,I}^{c}\left( k \right) \right> &=\frac{w\left( k \right)}{\sqrt{N_w}},k_{i-1}\leq k\leq k_i, \\
&=0,\mathrm{else}.
}
\end{equation}
For the bin method, the overlap inside a bin interval will be a constant if $w=1$ and that will vary with wave number $k$ if $w$=$\delta_{l}^{j,I}$. The overlap is numerically calculated in RLLM as
\begin{equation}\label{e-RLLM-overlap}
\left< \phi _{l,j,I}^{n}\mid \phi _{l,j,I}^{c}\left( k \right) \right> =\int{\varphi _{l,j,I}^{n}(r)\varphi _{l,j,I}^{c}(k,r)dr},
\end{equation}
where $\varphi _{l,j,I}^{n}(r)$ is the radial wave function obtained by RLLM and $\varphi _{l,j,I}^{c}(k,r)$ is the radial wave function of continuum state with wave number $k$ as defined in Eq. (\ref{e-continuum-rwf}). Hence, the overlap is a smooth function of the wave number $k$ for RLLM.

Numerov algorithm is a well-known solution for coupled channel equations. Yang \cite{Yang1980} provided an improved algorithm by using the iteration of the linear relationship between the radial wave functions at two neighbouring points. This method can well improve the computational stability, which benefits the integration of CDCC equations significantly as the number of channels in CDCC is typically larger than those in normal coupled channel calculations. Closed channels can also be treated conveniently. This algorithm is presented in \ref{appe-Yang}. A stabilization method is given in \ref{appe-Stable} to improve the linear dependence of the coupled channels solutions, which can be used to solve the CDCC equations including closed channels.

For all calculations in this paper, the step size $\Delta R$=0.05 fm is always adopted. It ensures the computational stability for all calculations.

\section{Benchmark calculation: $d$+$^{58}$Ni at EL=80.0 MeV}\label{sec-benchmark}

We firstly perform calculations for $d$+$^{58}$Ni reaction at EL=80.0 MeV, which has been variously studied in the past \cite{Austern1987,Yahiro1986,Matsumoto2003,Moro2009,Druet2010}. It would be a good test to check the correctness of our work. In the successful calculation by T. Druet et al. \cite{Druet2010}, the coupled channel equations were solved by $R$-matrix approach. In that way, it was required to compute the inversion of a $N_{ch}N_{var} \times N_{ch}N_{var}$ matrix, where $N_{ch}$ is the number of channels and $N_{var}$ is the number of basis functions used to solve coupled channel equations. As shown in Ref. \cite{Druet2010}, $N_{var}$ should be 50 and 100 to obtain converging elastic scattering angular distribution and breakup cross section respectively.

In Yang's improved Numerov algorithm \cite{Yang1980}, it is required to perform matrix multiplication and matrix inversion both for $R_m / \Delta R$ times (if $R_0$=0, see \ref{appe-Yang} for details). $R_m$ is the asymptotic radius. The dimensions of the matrices to be computed are all $N_{ch} \times N_{ch}$. For this reaction, $R_m$=30 fm and $\Delta R$=0.05 fm are overly enough to obtain converging results. As the time complexity of the inversion of a $M \times M$ matrix and the multiplication of two $M \times M$ matrices are both $O(M^3)$, Yang's method will be much faster than the $R$-matrix approach ( $R_m/\Delta R\ll N_{var}^{3}$ ) in solving coupled channel equations if $N_{ch}$ is same for two methods.

It should be emphasized that we perform calculations on this reaction just for verifying the correctness of our work. Therefore, we use the same potentials and calculation conditions as T. Druet et al.\cite{Druet2010} and compare our results with experimental data and their work. The ground state of the deuteron is restricted to be $1s$ state. $l_{\max}$=$\lambda_{\max}$=4. The spins of proton and neutron are ignored so that $l$=$j$=$I$. The continuum states up to $k_{\max}$=1.2 fm$^{-1}$ are taken into calculations.

\begin{figure}[tbp]
  \centering
  \includegraphics[width=0.6\textwidth]{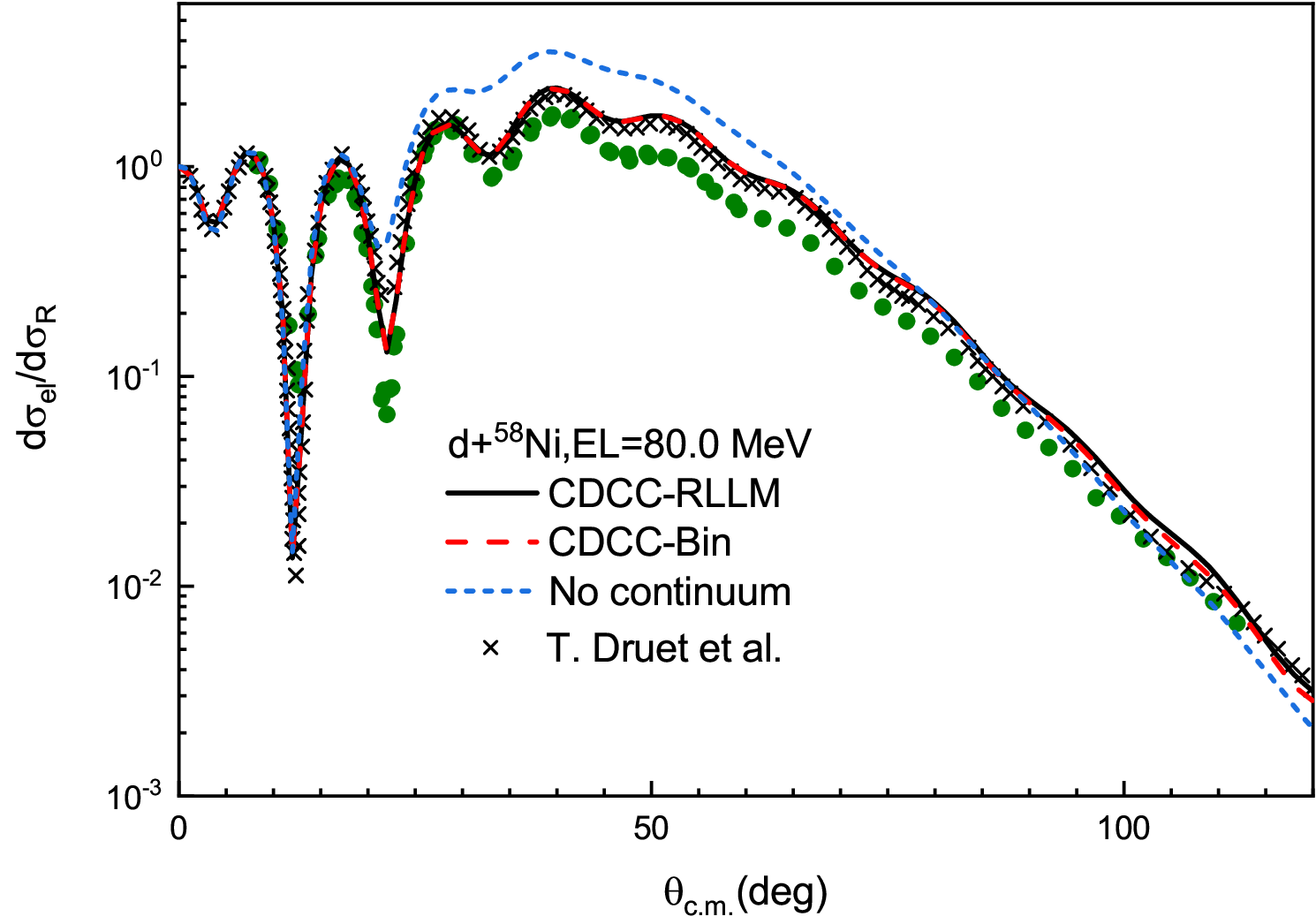}
  \caption{Calculated elastic scattering angular distributions in the Rutherford ratio for $d$+$^{58}$Ni reaction at EL=80.0 MeV. The solid and dashed lines represent the CDCC-RLLM and CDCC-Bin results respectively. The short dashed line denotes the calculated results without continuum channels. The circles are the experimental data taken from Ref \cite{Stephenson1983}. The crosses are the CDCC results calculated by T. Druet et al.\cite{Druet2010}. See text for details.}
  \label{fig-elastic-d+58Ni}
\end{figure}

For elastic scattering, $R_m$=15 fm. The CDCC calculation with RLLM (CDCC-RLLM) adopts $N$=30 and $h$=0.4 fm. In Ref. \cite{Druet2010}, the odd partial waves in the $p-n$ relative motion are neglected. In order to compare our results with those of T. Druet et al.\cite{Druet2010}, the odd partial waves are neglected in the present calculation and the comparison is shown in Fig. \ref{fig-elastic-d+58Ni}. The CDCC calculation with bin method (CDCC-Bin) adopts the width of bin $\Delta k$=$k_i - k_{i-1}$=0.04 fm$^{-1}$ and $r_{bin}$=40 fm. In this case, $N_{0,0,0}$=31 and $N_{2,2,2}=N_{4,4,4}$=30 for CDCC-Bin, while $N_{0,0,0}$=15, $N_{2,2,2}$=13 and $N_{4,4,4}$=12 for CDCC-RLLM. The results of the two methods and those calculated by T. Druet et al.\cite{Druet2010} are almost the same. Compared with the one-channel results which are calculated without continuum channels, CDCC calculation improves the results in the angles from 20° to 80°. The difference shows the sizable effect of the continuum coupling.

\begin{figure}[tbp]
  \centering
  \includegraphics[width=0.8\textwidth]{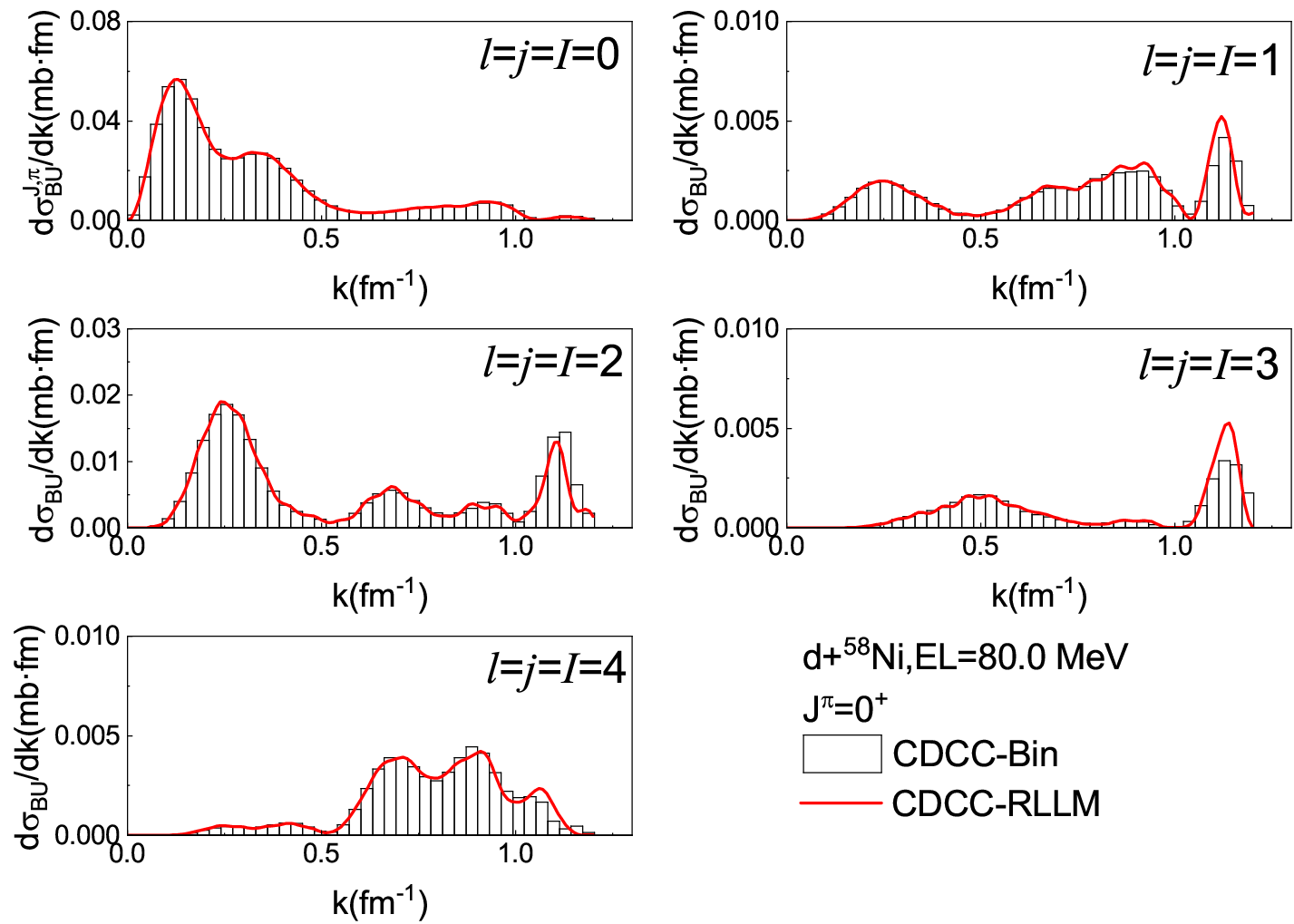}
  \caption{Calculated $d\sigma _{BU}^{J,\pi}/ dk$ for $d$+$^{58}$Ni reaction at EL=80.0 MeV at $J^{\pi}$=0$^+$. The histograms are the CDCC-Bin results and the solid lines represent the CDCC-RLLM results.}
  \label{fig-dbudk-d+58Ni-J=0}
\end{figure}

\begin{figure}[tbp]
  \centering
  \includegraphics[width=0.8\textwidth]{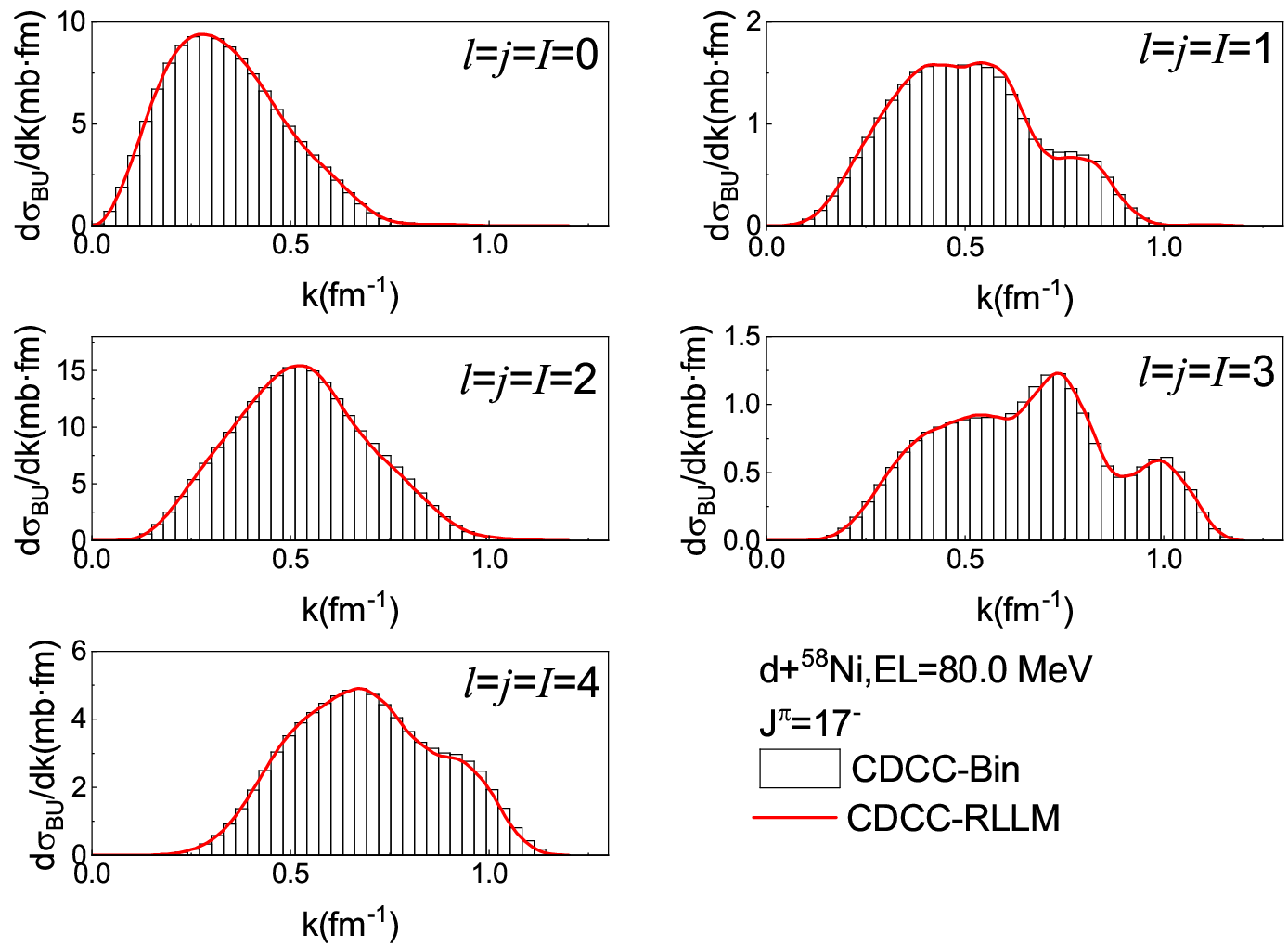}
  \caption{Same as Fig. \ref{fig-dbudk-d+58Ni-J=0} but at $J^{\pi}$=17$^-$.}
  \label{fig-dbudk-d+58Ni-J=17}
\end{figure}

The odd partial waves of the $p-n$ wave functions are included in breakup reaction calculations. $R_m$=27 fm. Finer discretization is done in both two methods to make the breakup reaction cross section converging. CDCC-Bin uses $\Delta k$=0.03 fm$^{-1}$ and $r_{bin}$=100 fm. CDCC-RLLM adopts $\{ N$=70, $h$=0.4 fm $\}$. We perform the calculations at $J^{\pi}$=0$^+$ and 17$^{-}$. The centrifugal term is minimum at $J^{\pi}$=0$^+$ and the partial-wave breakup reaction cross section is maximal at $J^{\pi}$=17$^-$. Hence, they are good examples to check our breakup calculations.  Satisfying agreement on $d\sigma _{BU}^{J,\pi} / dk$ is obtained for two discretization methods, as shown in Fig. \ref{fig-dbudk-d+58Ni-J=0} and \ref{fig-dbudk-d+58Ni-J=17}. The CDCC-Bin results are plotted in histograms as the $d\sigma _{BU}^{J,\pi} / dk$ calculated by bin method is a constant in each bin interval. In this case, the number of discretized states is 201 for CDCC-Bin, while that is 165 for CDCC-RLLM. CDCC-RLLM with less discretized states can provide the same $d\sigma _{BU}^{J,\pi} / dk$ as CDCC-Bin gives.

\begin{figure}[tbp]
  \centering
  \includegraphics[width=0.6\textwidth]{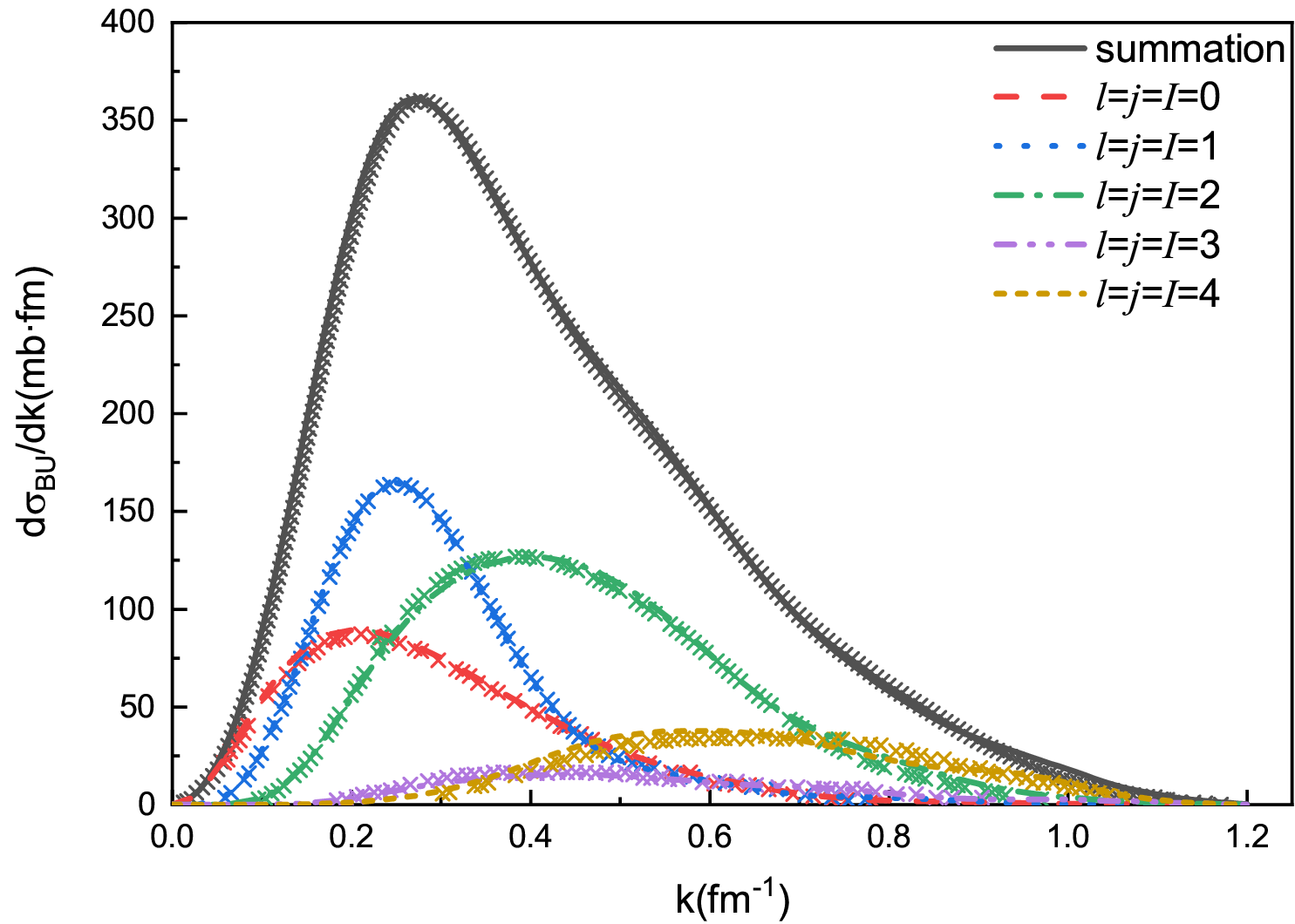}
  \caption{The differential breakup reaction cross section for $d$+$^{58}$Ni reaction at EL=80.0 MeV. The different types of curves represent the CDCC-RLLM results corresponding to different $p$-$n$ partial waves and the summation, while the crosses represent the calculated results by T. Druet et al.\cite{Druet2010}.}
  \label{fig-dbudk-d+58Ni-all}
\end{figure}

As there is no available breakup experimental data in this case, we compare the CDCC-RLLM results of $d\sigma _{BU} / dk$ with those taken from Ref. \cite{Druet2010} as shown in Fig. \ref{fig-dbudk-d+58Ni-all}. A satisfying agreement is obtained for each component corresponding to different $p-n$ partial waves and their summation.

The correctness of our calculations is verified by the above comparisons.

\section{Applications of RLLM in CDCC for $^6$Li induced reactions}\label{sec-appl}

\subsection{$\alpha +d$ model for $^6$Li}

$^6$Li is a typical weakly bound nucleus and its induced reactions have been of interest for both experimental and theoretical nuclear physicists for decades. The breakup effect is very important in the analysis of $^6$Li induced reactions. Different from the $d$ induced reactions, the reactions induced by $^6$Li are influenced severely by its resonance states. Therefore, $^6$Li induced reaction will be a good test to assess the capacity of RLLM in dealing with the effect of resonance states.

\begin{table}[tbp]
\caption{Parameters of $l$-dependent $\alpha-d$ interaction $V_{\alpha -d}^l$. $R_0$=2.1 fm and $a$=0.65 fm. }
\label{table-Vax}
\begin{indented}
\item[]
\begin{tabular}{cccc}
\br
 $l$ & 0 & 1 & 2  \\
\mr
$V_{0}^l$ (MeV)      & 67.69 &  63.90 & 65.33   \\
$V_{0}^{so,l}$ (MeV) &  0.00 &   5.72 &  4.78   \\
\br
\end{tabular}
\end{indented}
\end{table}

\begin{table}[tbp]
\caption{Calculated resonance energies $\varepsilon _{res} ^{cal}$ and resonance widths $\Gamma _{res} ^{cal}$ compared with the experimental value $\varepsilon _{res} ^{exp}$ and $\Gamma _{res} ^{exp}$\cite{Tilley2002a}.}
\label{table-res}
\begin{indented}
\item[]
\begin{tabular}{ccccc}
\br
  state & $\varepsilon _{res} ^{cal}$ & $\Gamma _{res} ^{cal}$ & $\varepsilon _{res} ^{exp}$ & $\Gamma _{res} ^{exp}$ \\
        &     (MeV)                   &        (MeV)           &           (MeV)             &      (MeV)             \\
\mr
 3$^+$   & 0.710 & 0.084 & 0.716 & 0.024  \\
 2$^+$   &  3.00 &  1.12 &  2.84 &  1.30  \\
 1$^+$   &  4.24 &  2.93 &  4.18 &  1.50  \\
\br
\end{tabular}
\end{indented}
\end{table}

\begin{figure}[tbp]
  \centering
  \includegraphics[width=0.6\textwidth]{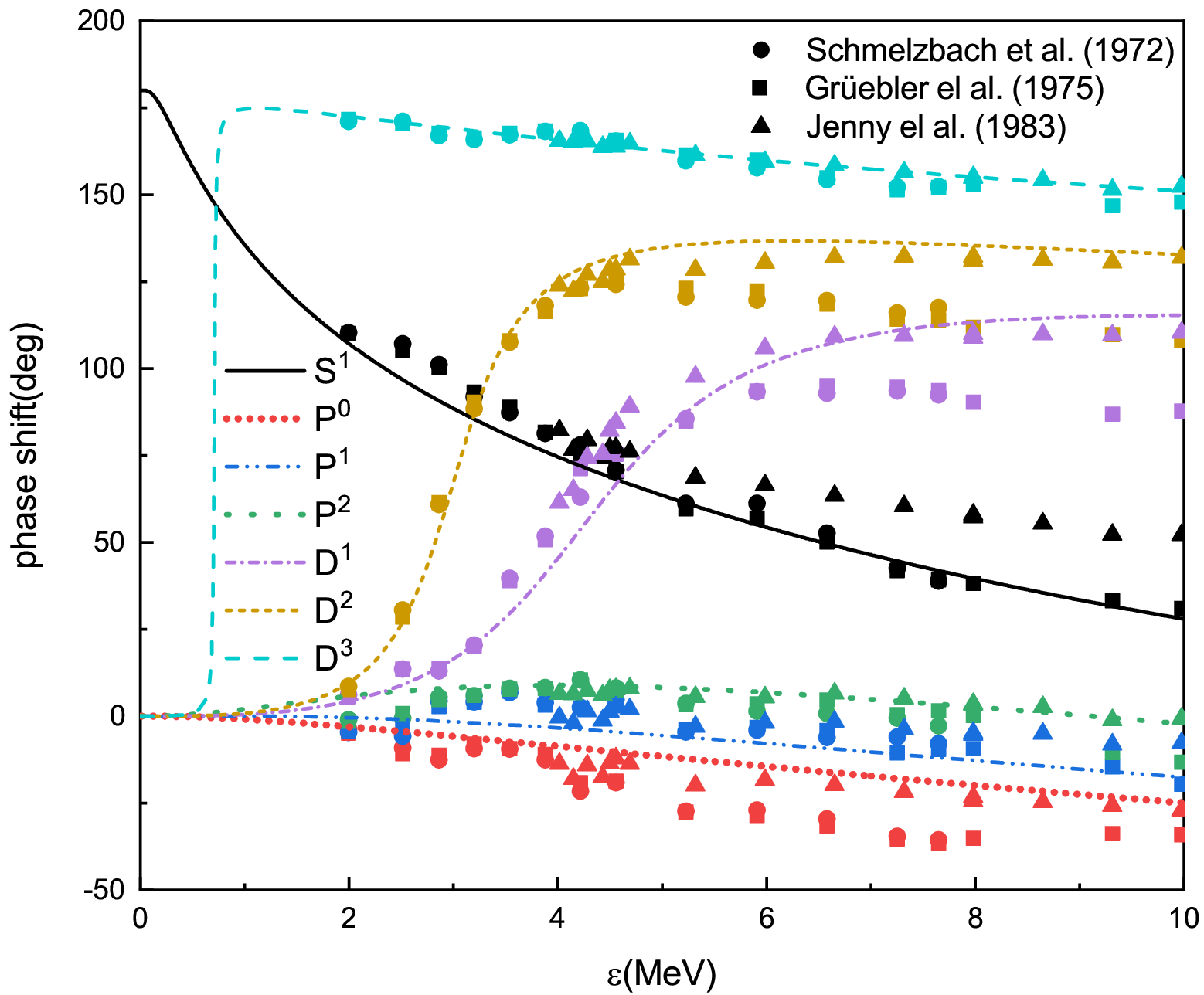}
  \caption{Calculated phase shift for $\alpha$-$d$ scattering. The experimental data are taken from Schmelzbach et al. \cite{Schmelzbach1972} (circles), Gr{\"{u}}ebler et al. \cite{Gruebler1975} (squares) and Jenny et al. \cite{Jenny1983} (triangles). $\varepsilon$ is the $\alpha$-$d$ relative energy in centre of mass system.}
  \label{f-ps-6Li}
\end{figure}

$\alpha $ and $d$ are regarded as the core and valence particles of $^6$Li respectively. As the spin of $\alpha$ is zero, $j$=$I$. The interaction between $\alpha$ and $d$ is chosen to be $l$-dependent and in the Woods-Saxon form.
\begin{eqnarray}\label{e-WSpot}
V_{\alpha -d}^{l} &=-V_{0}^{l}f( r ) +V_{0}^{so,l}\frac{\lambda _{\pi}^{2}}{r}\frac{d}{dr}f( r )  ( \bi{l} \cdot \bi{s}  ) +V_C, \\
f ( r  ) &=\frac{1}{1+\exp \left\{ \left( r-R_0 \right) /a \right\}} \\
V_C &=Z_{\alpha}Z_{d}e^2/r,r \geq R_0 \\
&=\frac{Z_{\alpha}Z_{d}e^2}{2R_0}\left( 3-\frac{r^2}{R_{0}^{2}} \right) ,r < R_0
\end{eqnarray}
where $\lambda _{\pi}^{2}$=2.00 fm$^2$. $Z_{\alpha}$ and $Z_d$ are the charge numbers of $\alpha$ and $d$ respectively. The parameters of $V_{\alpha -d}^l$ for $l$=0, 1 and 2 are listed in Table \ref{table-Vax}. $V_{\alpha -d}^l$ can not only generate the binding energy for $^{6}$Li correctly (-1.47 MeV), but also can well reproduce the resonance energies and resonance widths for $^{6}$Li resonance states as shown in Table \ref{table-res}. Comparison between calculated phase shifts and experimental data \cite{Schmelzbach1972,Gruebler1975,Jenny1983} is shown in Fig. \ref{f-ps-6Li}. In general, the experimental data of Jenny et al. \cite{Jenny1983} are consistent with those of Schmelzbach et al. \cite{Schmelzbach1972} and Gr{\"{u}}ebler et al. \cite{Gruebler1975} below 5 MeV and the calculated phase shifts match the three groups of experimental data well in this energy region. However, the data of Jenny et al. \cite{Jenny1983} are visibly larger than those of Schmelzbach et al. \cite{Schmelzbach1972} and Gr{\"{u}}ebler et al. \cite{Gruebler1975} for $S^1$, $P^{0,1,2}$ and $D^{1,2}$ partial waves in 5-10 MeV. In this energy region, our calculated phase shifts match the results of Jenny et al. \cite{Jenny1983} for $P$ and $D$ partial waves and are in consistent with the data of Schmelzbach et al. \cite{Schmelzbach1972} and Gr{\"{u}}ebler et al. \cite{Gruebler1975} for $S$ partial wave. The $V_{\alpha -d}^l$ for $l$>0 is assumed to be parity dependent. $V_{\alpha -d}^1$ and $V_{\alpha -d}^2$ are adopted to calculate $\alpha-d$ wave functions of higher odd and even partial waves respectively.

\begin{figure}[tbp]
  \centering
  \includegraphics[width=0.6\textwidth]{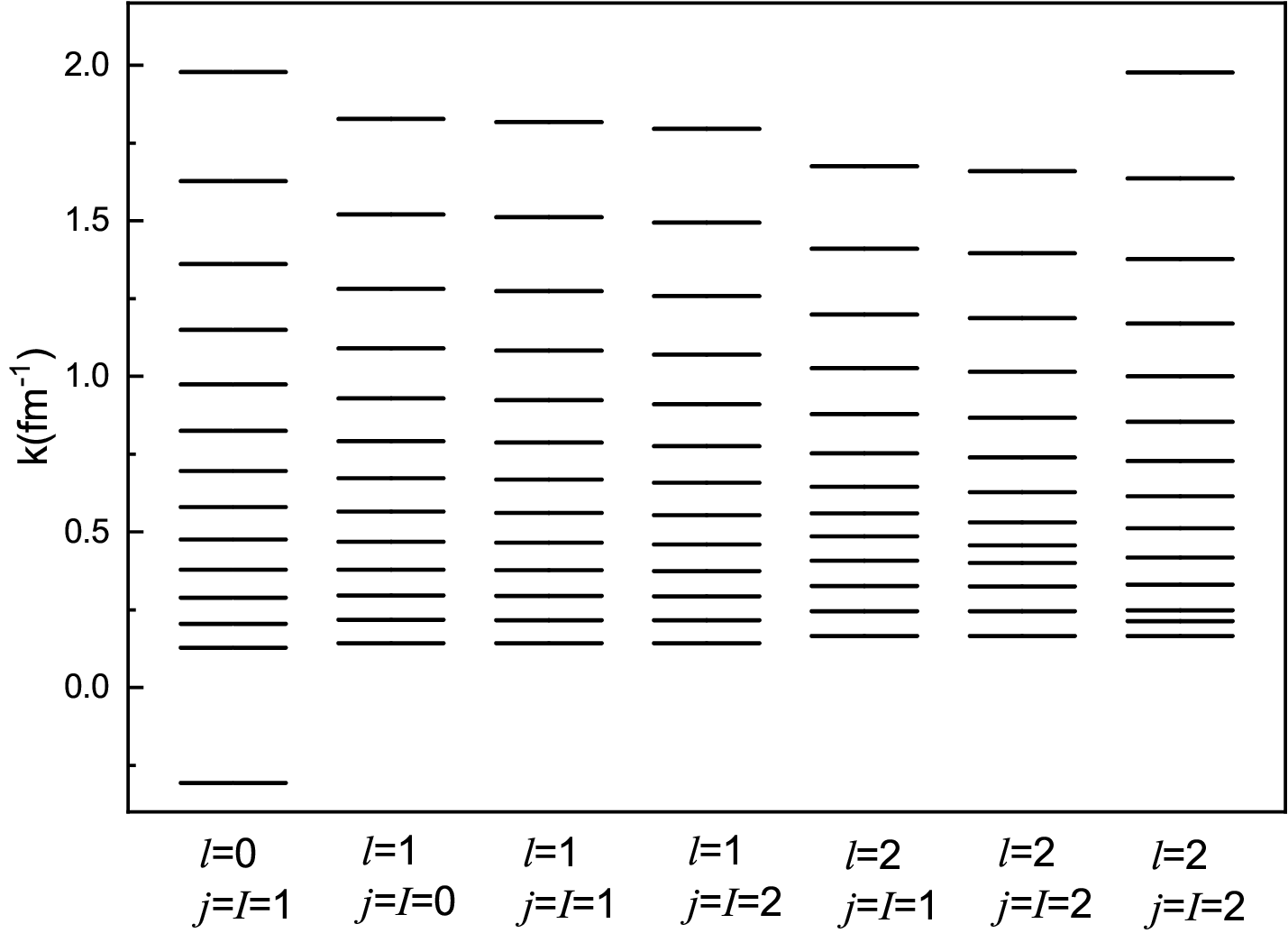}
  \caption{Discrete momentum of $\phi _{l,j,I}^{n}$ generated by RLLM for $^6$Li with RLLM parameters $\{N$=20,$h$=0.5 fm $\}$ up to $l_{\max}=2$ and $k_{\max}$=2.0 fm$^{-1}$. The line with negative value corresponds to the ground state of $^6$Li.}
  \label{fig-k-6Li}
\end{figure}

$h$ is optimized to be 0.5 fm for $V_{\alpha-d}$. We only vary $N$ to change the $^6$Li discretized states. Fig. \ref{fig-k-6Li} shows the pseudo state momentums up to $l_{\max}=2$ and $k_{\max}$=2.0 fm$^{-1}$ with RLLM parameters $N$=20 and $h$=0.5 fm. For each calculation in Sec. \ref{appl-6+12} and \ref{appl-6+59}, $\lambda_{\max}$ is set to be 2$l_{\max}$ to include all couplings.

\subsection{$^6$Li+$^{12}$C at EL=168.6 and 178.0 MeV}\label{appl-6+12}

As there is no elastic scattering and breakup experimental data for $^6$Li+$^{12}$C reaction at the same incident energy, we calculate the elastic scattering angular distribution and breakup reaction cross section at EL=168.6 MeV and 178.0 MeV respectively. For the $\alpha$-$^{12}$C optical potential, we adopt the optical potential at $\alpha$ incident energy of 120 MeV in Ref. \cite{Amer2022}, which is based on the SPP2 effective $NN$ interaction \cite{Chamon2021}. Its parameters are listed in the Table 5 of Ref. \cite{Amer2022}. The $d$-$^{12}$C optical potential is taken from Ref. \cite{Zhang2016}.

\begin{table}[tbp]
\caption{The elastic scattering $S$-matrix $S^{J,\pi}_{\beta_0,\beta_0}$ (multiplied by 1000) of $^6$Li+$^{12}$C reaction at EL=168.6 MeV as a function of the maximum momentum for $^6$Li continuum $k_{\max}$ (top, $l_{\max}$=4, $N$=20, $R_m$=18 fm), of the number of RLLM basis $N$ (second top, $l_{\max}$=4, $k_{\max}$=2.2 fm$^{-1}$, $R_m$=18 fm), of asymptotic radius $R_m$ (third top, $l_{\max}$=4, $N$=20, $k_{\max}$=2.2 fm$^{-1}$) and of maximum angular momentum for $^6$Li continuum $l_{\max}$ (bottom, $N$=20, $R_m$=18 fm, $k_{\max}$=2.2 fm$^{-1}$). }
\label{table-elas-smat-6+12}
\begin{indented}
\item[]
\begin{tabular}{ccccc}
\br
  $k_{\max}$ (fm$^{-1}$) & 1.6 & 1.8 & 2.0 & 2.2  \\
\mr
 $J^{\pi}$=0$^{-}$  &  -0.08-4.86$i$ &  -0.06-4.42$i$ &  -0.19-4.35$i$ &  -0.16-4.33$i$ \\
 $J^{\pi}$=12$^{+}$ &  -12.6+4.31$i$ &  -12.3+4.19$i$ &  -12.2+4.18$i$ &  -12.2+4.18$i$ \\
 $J^{\pi}$=27$^{-}$ & 626.0+254.1$i$ & 626.2+254.0$i$ & 626.2+254.1$i$ & 626.2+254.1$i$ \\
\br
 $N$ & 10 & 15 & 20 & 25 \\
\mr
 $J^{\pi}$=0$^{-}$  &  -0.26-4.30$i$ &  -0.16-4.33$i$ &  -0.16-4.33$i$ &  -0.16-4.33$i$ \\
 $J^{\pi}$=12$^{+}$ &  -12.2+4.53$i$ &  -12.2+4.21$i$ &  -12.2+4.18$i$ &  -12.2+4.17$i$ \\
 $J^{\pi}$=27$^{-}$ & 624.0+254.8$i$ & 625.8+254.3$i$ & 626.2+254.1$i$ & 626.3+254.1$i$ \\
\br
 $R_m$ (fm) & 9 & 12 & 15 & 18 \\
\mr
 $J^{\pi}$=0$^{-}$  &  -0.17-4.36$i$ &  -0.18-4.32$i$ &  -0.16-4.32$i$ &  -0.16-4.33$i$ \\
 $J^{\pi}$=12$^{+}$ &  -12.5+4.03$i$ &  -12.1+4.15$i$ &  -12.2+4.16$i$ &  -12.2+4.18$i$ \\
 $J^{\pi}$=27$^{-}$ & 628.9+253.5$i$ & 626.5+254.0$i$ & 626.3+254.1$i$ & 626.2+254.1$i$ \\
\br
 $l_{\max}$ & 0 & 2 & 4 & 6 \\
\mr
 $J^{\pi}$=0$^{-}$  &  -0.24-2.11$i$  &  -0.51-4.53$i$ &  -0.16-4.33$i$ &  -0.40-4.15$i$ \\
 $J^{\pi}$=12$^{+}$ &  -7.60+7.10$i$  &  -12.4+35.1$i$ &  -12.2+4.18$i$ &  -11.7+4.36$i$ \\
 $J^{\pi}$=27$^{-}$ & 582.6+337.3$i$  & 621.1+259.6$i$ & 626.2+254.1$i$ & 627.2+254.7$i$ \\
\br
\end{tabular}
\end{indented}
\end{table}

It is crucial to choose appropriate $k_{\max}$, $N$, $R_m$ and $l_{\max}$ to perform CDCC calculations. For elastic scattering, we only need to focus on the convergence of elastic scattering $S$-matrix $S^{J,\pi}_{\beta_0,\beta_0}$. Table \ref{table-elas-smat-6+12} shows some $S^{J,\pi}_{\beta_0,\beta_0}$ with different parameters. The elastic scattering $S$-matrix $S^{J,\pi}_{\beta_0,\beta_0}$ converges fast as $k_{\max}$, $N$ and $R_m$ increase, while it converges slowly as $l_{\max}$ increases. The difference between the $S^{J,\pi}_{\beta_0,\beta_0}$ calculated with $l_{\max}$=4 and $l_{\max}$=6 is little for $J^{\pi}$=27$^{-}$ while that is over 5$\%$ for $J^{\pi}$=0$^{-}$ and 12$^{+}$.

\begin{figure}[tbp]
  \centering
  \includegraphics[width=0.6\textwidth]{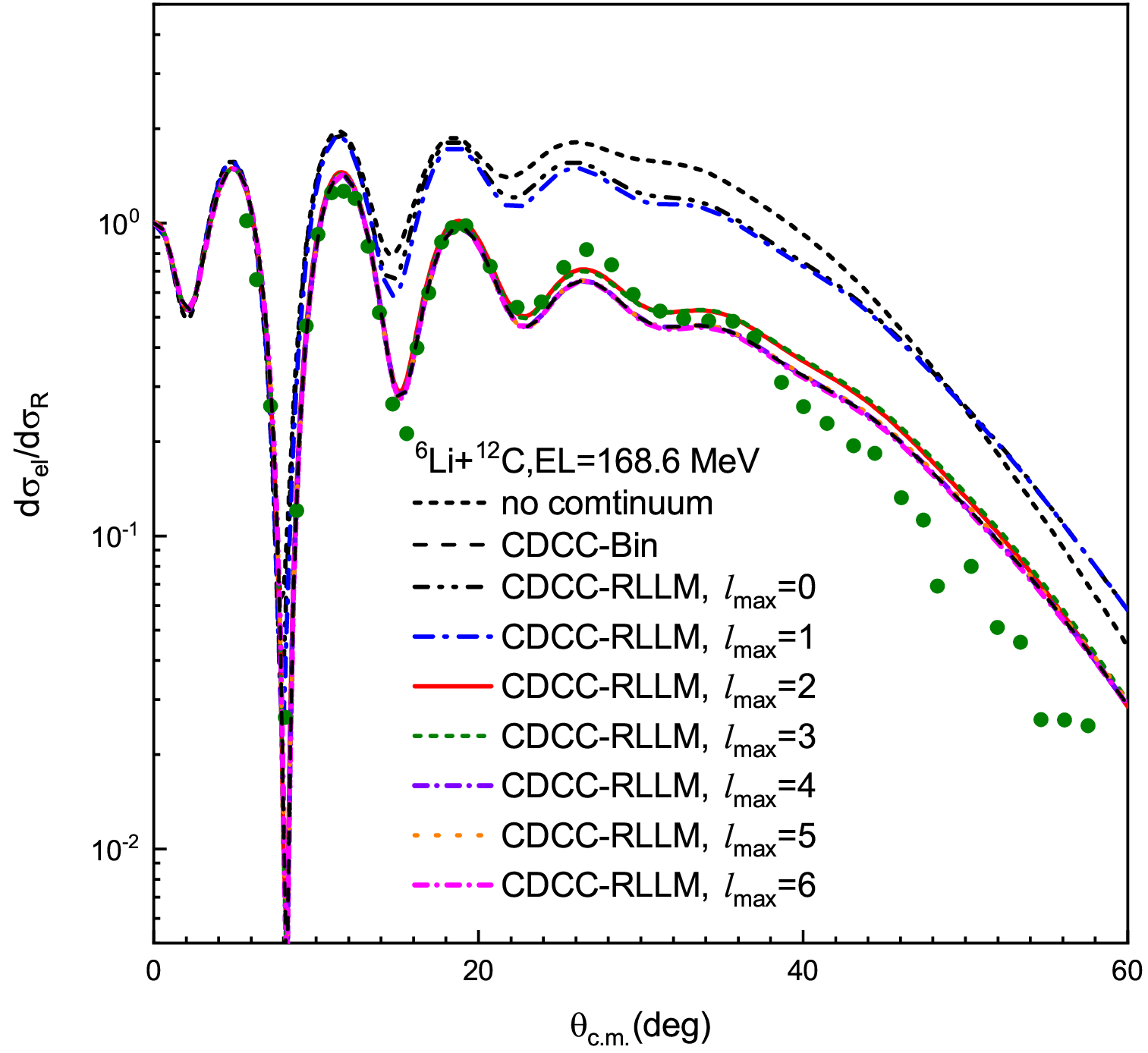}
  \caption{Calculated elastic scattering angular distributions in the Rutherford ratio for $^6$Li+$^{12}$C at EL=168.6 MeV. The circles are the experimental data taken from Ref \cite{Katori1988}. The CDCC-RLLM results with $l_{\max}$=2 and 3 are almost the same. The CDCC-RLLM results with $l_{\max}$=4, 5, 6 and the CDCC-Bin results are almost the same. See text for details.}
  \label{fig-elastic-6+12}
\end{figure}

In order to investigate the coupling effect of the continuum with high orbital angular momentum, we calculate the elastic scattering angular distributions of $^6$Li+$^{12}$C at EL=168.6 MeV with progressively increasing $l_{\max}$. $N$=20, $k_{\max}$=2.2 fm$^{-1}$ and $R_m$=18 fm are adopted for calculations and the results are presented in Fig. \ref{fig-elastic-6+12}. The elastic scattering angular distribution converges with $l_{\max}$=4, which suggests that the coupling effect of $l$>4 continuum states are negligible on elastic scattering for this reaction. On the other hand, the elastic scattering angular distribution calculated with $l_{\max}$=2 is slightly larger than that calculated with $l_{\max}$=4 from 20 to 50 degrees and they are both in reasonable agreement with the experimental data \cite{Katori1988}. Moreover, it is observed that the one-channel calculated results, whose continuum channels are omitted, are similar to the CDCC-RLLM results with $l_{\max}$=0 and 1, while they are notably different from the CDCC results with $l_{\max} \geq$2. The discrepancy shows that the inclusion of the coupling with continuum, especially the $D$-wave continuum, can significantly improve the agreement with the experimental data.

\begin{table}[tbp]
\caption{Bin scheme of $^6$Li $D$-wave continuum used for $^6$Li+$^{12}$C reaction calculations at EL=168.6 MeV and 178.0 MeV. See text for details.}
\label{table-bin-6+12}
\begin{indented}
\item[]
\begin{tabular}{ccc}
\br
  state &  momentum region (fm$^{-1}$) & $\Delta k$ (fm$^{-1}$)  \\
\mr
 $l$=2, $j$=$I$=3    & 0$\leq k \leq$0.2 & 0.05  \\
 $l$=2, $j$=$I$=3    & 0.2$\leq k \leq$0.25 & 0.025  \\
 $l$=2, $j$=$I$=3    & 0.25$\leq k \leq$2.2 & 0.05  \\
 $l$=2, $j$=$I$=2    & 0$\leq k \leq$0.4 & 0.05  \\
 $l$=2, $j$=$I$=2    & 0.4$\leq k \leq$0.5 & 0.025  \\
 $l$=2, $j$=$I$=2    & 0.5$\leq k \leq$2.2 & 0.05  \\
 $l$=2, $j$=$I$=1    & 0$\leq k \leq$0.4 & 0.05  \\
 $l$=2, $j$=$I$=1    & 0.4$\leq k \leq$0.6 & 0.025  \\
 $l$=2, $j$=$I$=1    & 0.6$\leq k \leq$2.2 & 0.05  \\
\br
\end{tabular}
\end{indented}
\end{table}

For completeness, the elastic scattering angular distribution calculated by CDCC-Bin with $l_{\max}$=4 and $k_{\max}$=2.2 fm$^{-1}$ is also presented in Fig. \ref{fig-elastic-6+12}. The widths of bin interval $\Delta k$ for $l$=0, 1, 3 and 4 continuum states are all set to be 0.1 fm$^{-1}$. As the 3$^+$, 2$^+$ and 1$^+$ resonance states are located in the momentum regions 0.2 fm$^{-1}$ $\leq k \leq$ 0.25 fm$^{-1}$, 0.4 fm$^{-1}$ $\leq k \leq$ 0.5 fm$^{-1}$ and 0.4 fm$^{-1}$ $\leq k \leq$ 0.6 fm$^{-1}$ respectively, a finer bin scheme is required for $l$=2 continuum states as shown in Table. \ref{table-bin-6+12}. $w$=$\sin \delta _{l}^{j,I}$ for resonance bins and $w$=1 for non-resonance bins. With $R_m$=18 fm and $r_{bin}$=50 fm, this bin scheme can provide convergent elastic scattering angular distribution and breakup cross section for $^6$Li induced reactions at incident energies well above the Coulomb barrier. In this CDCC model space, the number of states is 174 for CDCC-RLLM ($N$=20, $l_{\max}$=4) and that is 360 for CDCC-Bin. The calculated elastic scattering angular distribution by CDCC-Bin is almost the same as the converging CDCC-RLLM result.

\begin{figure}[tbp]
  \centering
  \includegraphics[width=0.8\textwidth]{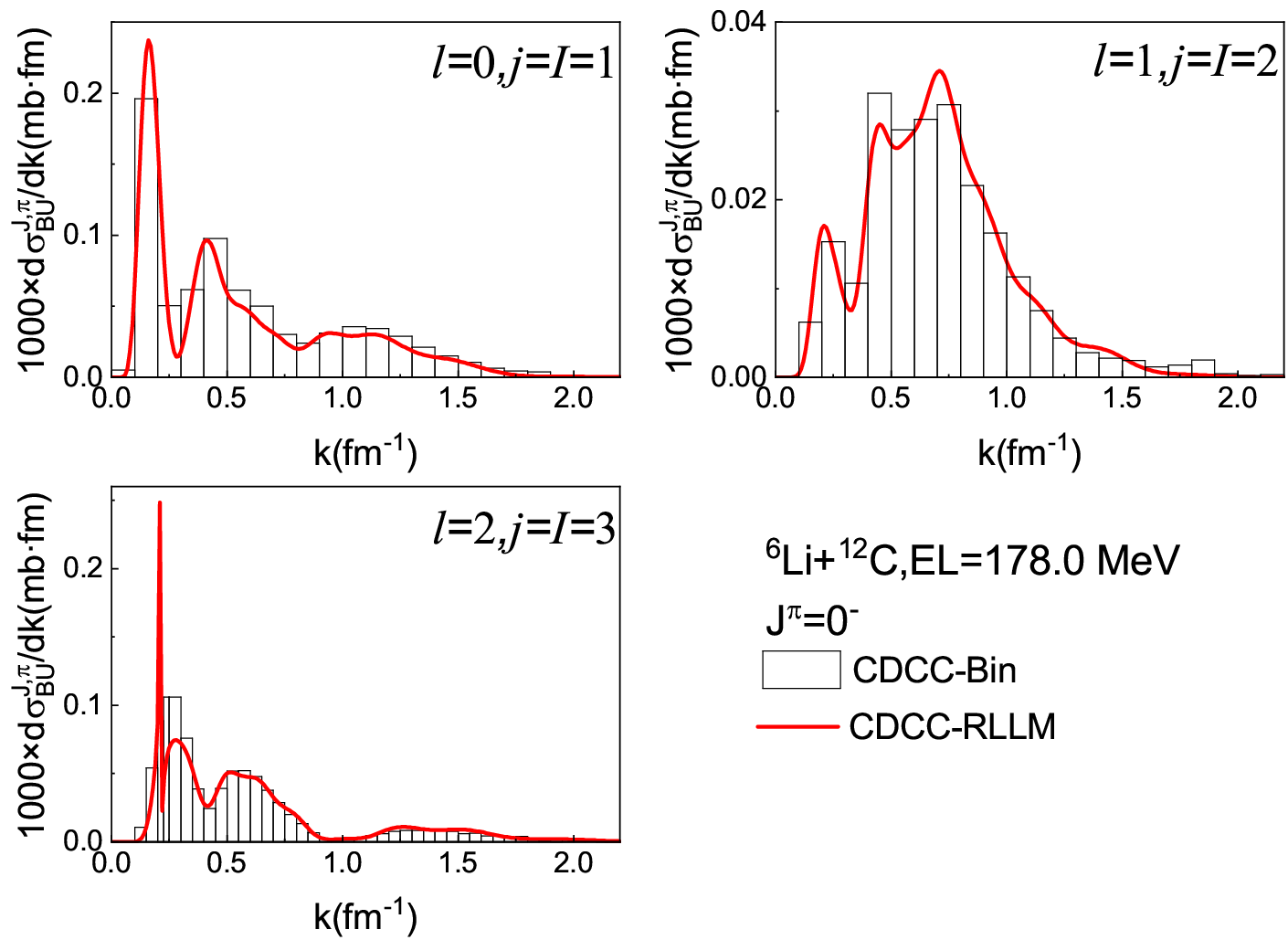}
  \caption{Same as Fig. \ref{fig-dbudk-d+58Ni-J=0} but for $^6$Li+$^{12}$C reaction at EL=178.0 MeV and $J^{\pi}$=0$^-$. The results are multiplied by 1000.}
  \label{fig-dbudk-6+12-J=0}
\end{figure}

\begin{figure}[tbp]
  \centering
  \includegraphics[width=0.8\textwidth]{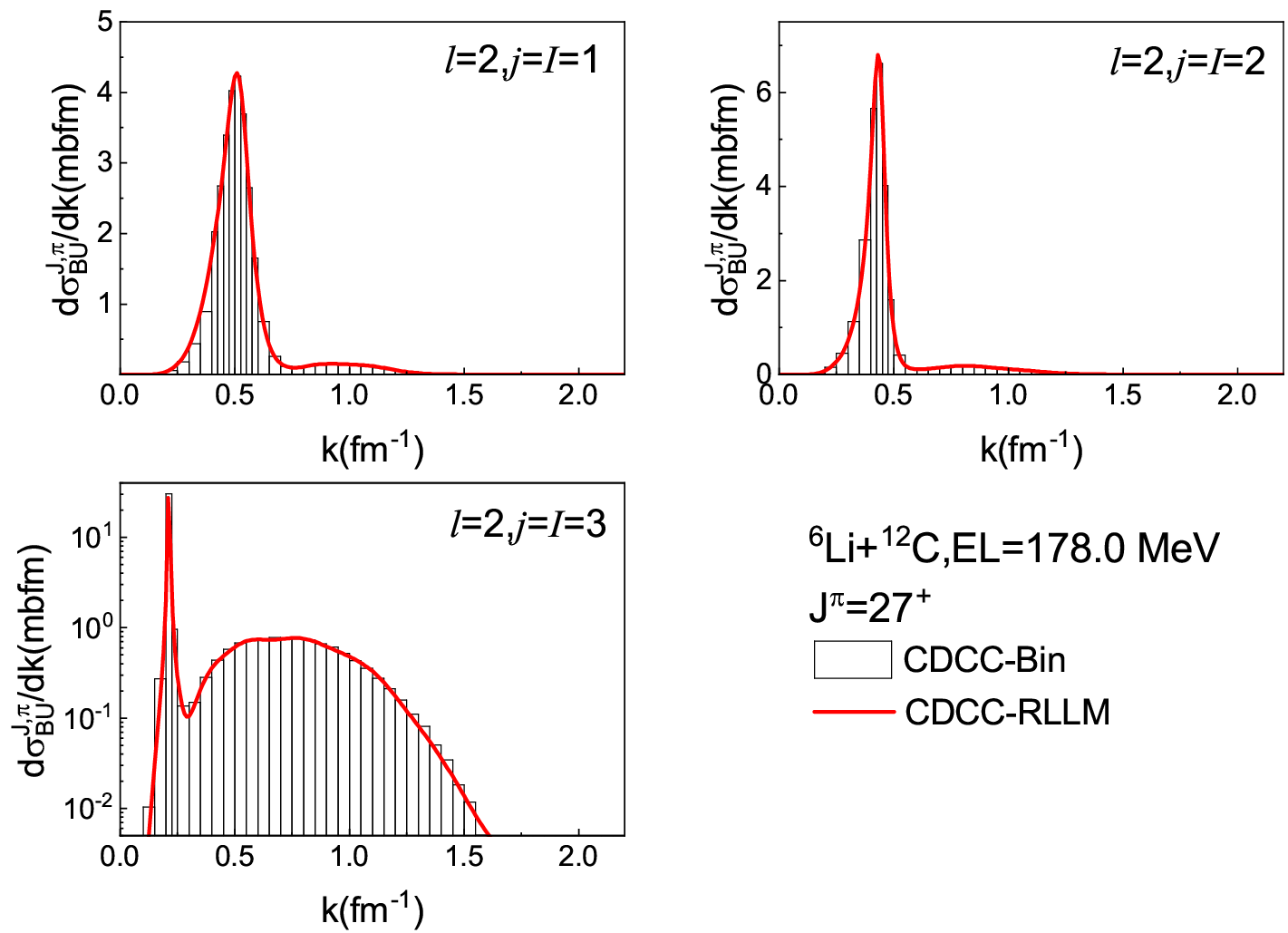}
  \caption{Same as Fig. \ref{fig-dbudk-d+58Ni-J=0} but for $^6$Li+$^{12}$C reaction at EL=178.0 MeV and $J^{\pi}$=27$^+$.}
  \label{fig-dbudk-6+12-J=27}
\end{figure}

For $^6$Li breakup reaction on $^{12}$C target at EL=178.0 MeV, $N$=20, $k_{\max}$=2.2 fm$^{-1}$ and $R_m$=18 fm are sufficient for CDCC-RLLM calculation to achieve convergence with different $l_{\max}$. We perform CDCC-Bin calculation with $l_{\max}$=4 for comparison. Fig. \ref{fig-dbudk-6+12-J=0} and \ref{fig-dbudk-6+12-J=27} show the $d\sigma _{BU}^{J,\pi} / dk$ of some $\alpha -d$ partial waves at $J^{\pi}$=0$^-$ and 27$^+$ respectively. It should be noted that the width of bin $\Delta k$ for resonance states is so small that the CDCC-Bin results are nearly a constant in any resonance bin. Therefore, we plot them in histograms for convenience. Good agreement is obtained for each case. The peaks for the $l=2$ continuum states, which arise from $^6$Li resonance states, are well described by CDCC-RLLM. Therefore, RLLM is an efficient approach to describe the resonant states and their effect on scattering.

\begin{figure}[tbp]
  \centering
  \includegraphics[width=0.6\textwidth]{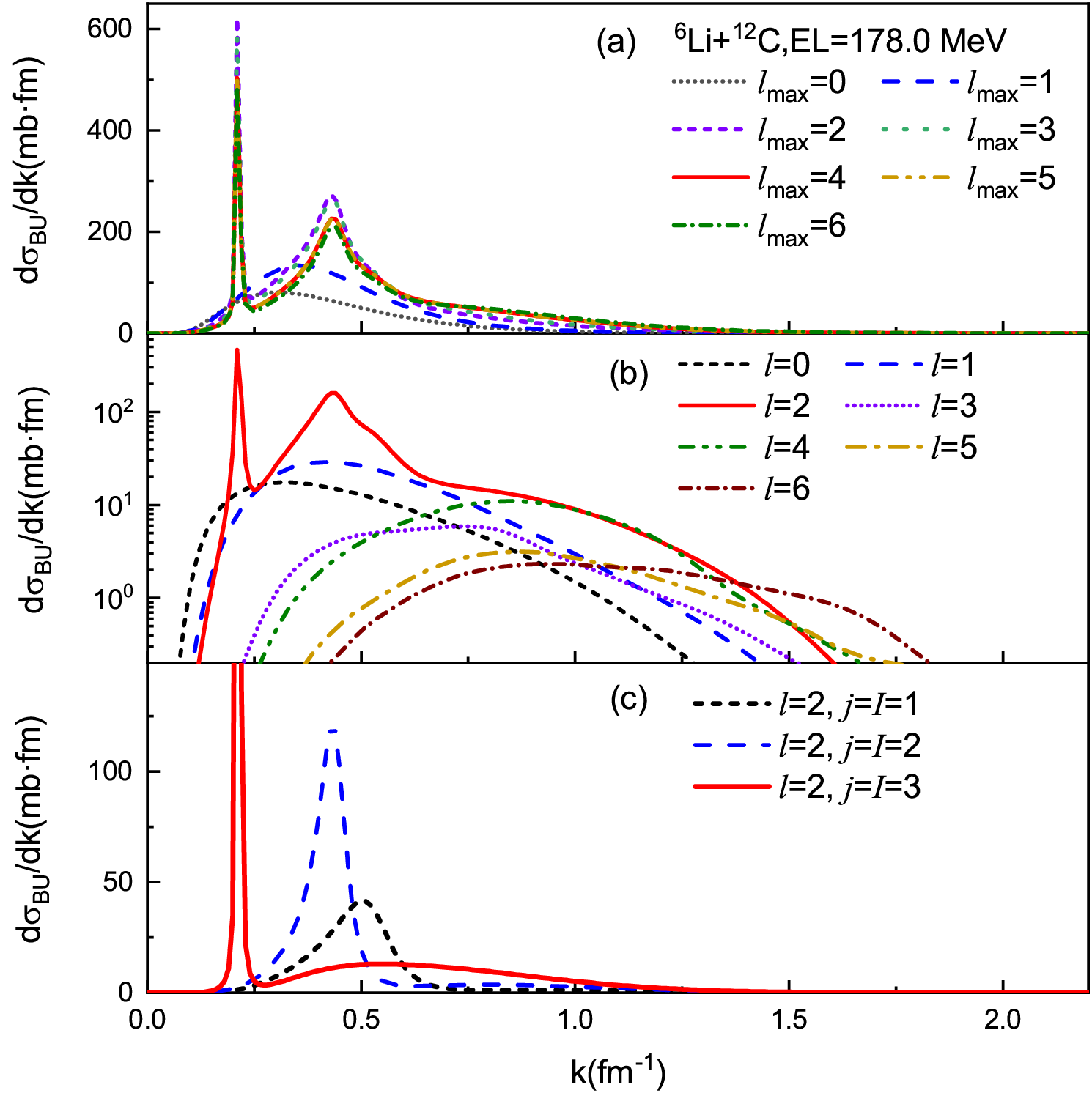}
  \caption{(a)Calculated differential breakup reaction cross sections with different $l_{\max}$ for $^6$Li+$^{12}$C reaction at EL=178.0 MeV. (b) The components of differential breakup reaction cross section calculated with $l_{\max}$=6. The individual components correspond to different $l$ states. (c) Differential breakup reaction cross sections calculated with $l_{\max}$=6 for $l$=2 states with $j$=$I$=1, 2 and 3 respectively.}
  \label{fig-dbudk-6+12-all}
\end{figure}

In Fig. \ref{fig-dbudk-6+12-all}, the differential breakup reaction cross sections calculated with different $l_{\max}$ are presented. The discrepancy between the curves is not appreciable when $l_{\max} \geq$4. The curve shape of differential breakup reaction cross section is mainly determined by the $l$=2 components. The peaks located at $k$=0.21 fm$^{-1}$, 0.43 fm$^{-1}$ and 0.50 fm$^{-1}$ in Fig. \ref{fig-dbudk-6+12-all}(c) correspond to the 3$^{+}$, 2$^{+}$ and 1$^{+}$ resonance states respectively. The main components of breakup reaction cross section come from the continuums with $l$=0, 1 and 2, while the contribution from the continuum with $l$=4 is not negligible around $k$=0.9 fm$^{-1}$.

\begin{figure}[tbp]
  \centering
  \includegraphics[width=0.6\textwidth]{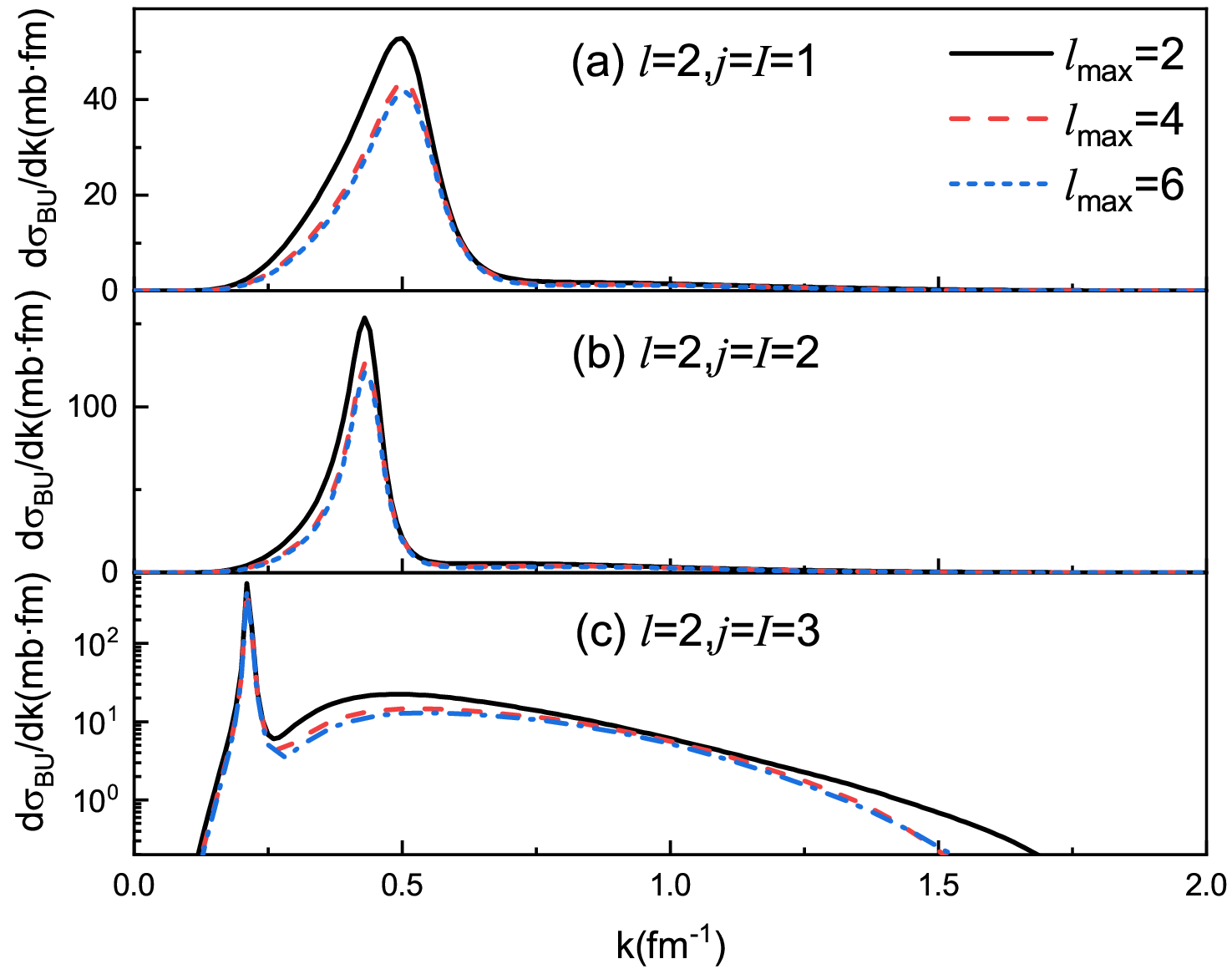}
  \caption{$l$=2 components of the differential breakup reaction cross sections for $^6$Li+$^{12}$C reaction at EL=178.0 MeV. (a) $l$=2, $j$=$I$=1. (b) $l$=2, $j$=$I$=2. (c) $l$=2, $j$=$I$=3. The solid, dashed and short dashed lines represent the results calculated with $l_{\max}$=2, 4 and 6 respectively.}
  \label{fig-dbudk-6+12-supp}
\end{figure}

Moreover, the suppression effect on breakup reaction cross section from the continuum with $l $>2 should be considered. The breakup reaction cross sections calculated with $l_{\max}$=2, 4 and 6 are 88.93 mb, 84.12 mb and 81.59 mb respectively. It is approximately reduced by 10$\%$ when $l$=3, 4, 5 and 6 continuum states are included in CDCC calculations. It can be seen in Fig. \ref{fig-dbudk-6+12-all}(a) that the suppression mainly occurs at 0.25 fm$^{-1}$ < $k$ < 0.55 fm$^{-1}$. Especially, the contributions from the resonance states are suppressed visibly as shown in Fig. \ref{fig-dbudk-6+12-supp}.

\begin{figure}[tbp]
  \centering
  \includegraphics[width=0.6\textwidth]{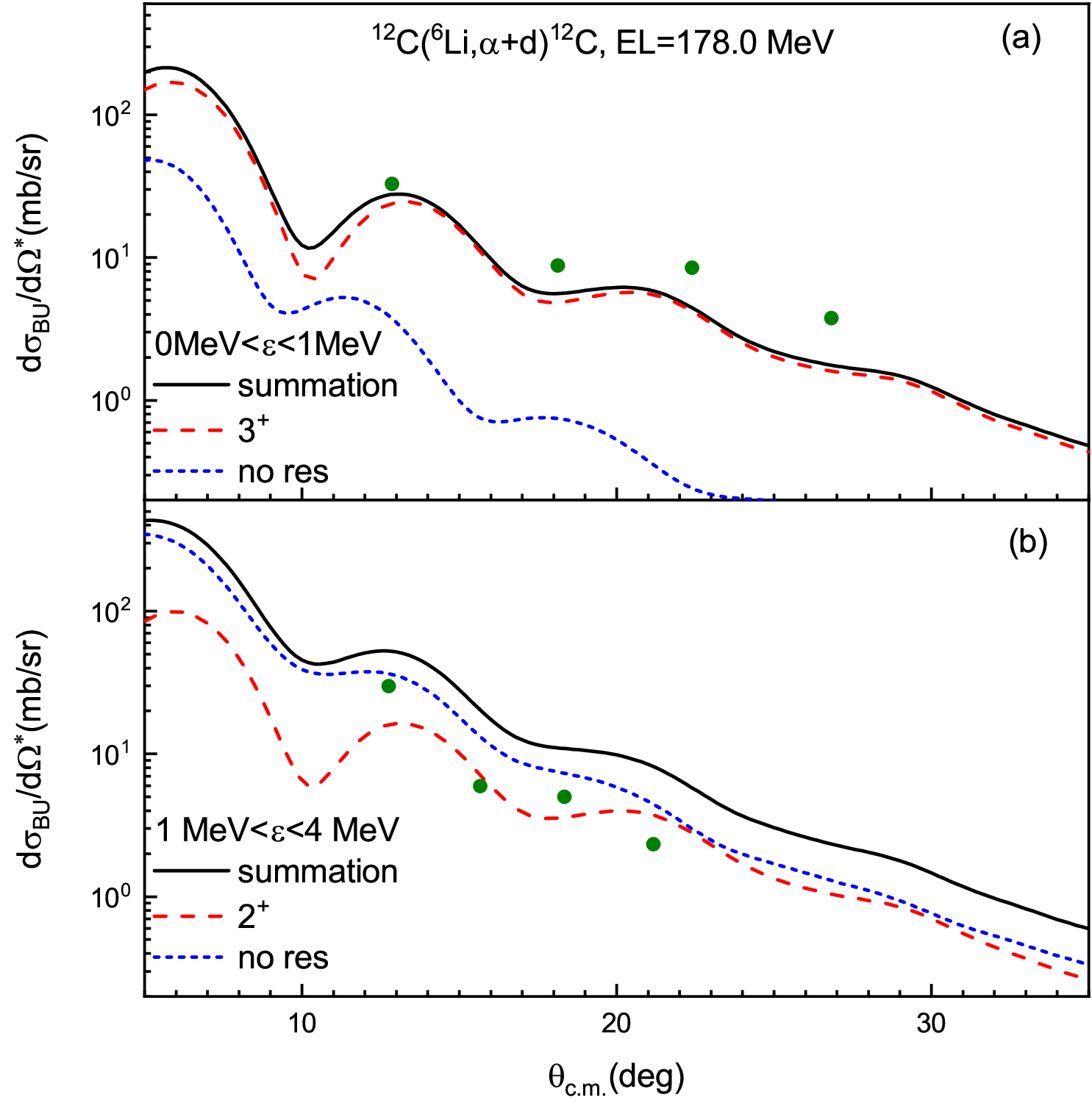}
  \caption{Calculated exclusive elastic breakup reaction cross sections for $^{12}$C($^6$Li,$\alpha +d$)$^{12}$C reaction at EL=178.0 MeV with respect to the scattering angle of the centre of mass of the $\alpha-d$ system. (a) The results for the breakup from the $\alpha-d$ continuum at 0.0 < $\varepsilon$< 1.0 MeV (energy region I,$\varepsilon _{I}$). (b) Same as (a) but for the breakup from the $\alpha-d$ continuum at 1.0 < $\varepsilon$< 4.0 MeV (energy region II,$\varepsilon _{II}$). The dashed and short dashed lines represent the contributions from resonance states (3$^+$ resonance state for $\varepsilon _{I}$ and 2$^+$ resonance state for $\varepsilon _{II}$) and non-resonant continuum states respectively. The solid lines represent the summations. The circles are the experimental data taken from Ref. \cite{Sakuragi1986}. See text for details.}
  \label{fig-dbudO-6+12}
\end{figure}

The exclusive cross section has been measured for elastic breakup reaction $^{12}$C($^6$Li,$\alpha +d$)$^{12}$C reaction at EL=178.0 MeV \cite{Sakuragi1986}. The cross sections ($d\sigma _{BU}/d\Omega^{*}$, where $\Omega^{*}$ represents the motion direction for the centre of mass of the $\alpha-d$ system) were given for the breakup from the continuum at 0.0<$\varepsilon$<1.0 MeV (energy region I,$\varepsilon _{I}$) and from the continuum at 1.0<$\varepsilon$<4.0 MeV (energy region II,$\varepsilon _{II}$), in which the 3$^+$ and 2$^+$ resonance states are located respectively.

Fig. \ref{fig-dbudO-6+12} shows the calculated results compared with the experimental data \cite{Sakuragi1986}. Since the calculated results with $l_{\max}$=4 and 6 are almost the same, $l_{\max}$=4 is used in the calculation. Reasonable agreements are obtained for two energy regions. The contribution from 3$^+$ resonance state is the main part of the breakup reaction cross section in $\varepsilon _{I}$, while the contributions from 2$^+$ resonance state and the non-resonant continuum states are both important in $\varepsilon _{II}$.

Moderate effects are found on elastic scattering angular distribution and breakup reaction cross section when $l$=3 and 4 continuum states are included in CDCC calculations for the reactions $^6$Li+$^{12}$C at EL=168.6 and 178.0 MeV.

\subsection{$^6$Li+$^{59}$Co at EL=12.0, 17.4 and 18.0 MeV}\label{appl-6+59}

We finally perform calculations on $^6$Li+$^{59}$Co reactions at EL=12.0, 17.4 and 18.0 MeV, which are close to the Coulomb barrier ($V_B$=12.0 MeV in center of mass system\cite{Beck2003}). The nuclear and Coulomb couplings with the continuum are both very important in this energy region. Moreover, the important role of closed channels is a well-known issue in the collisions in the vicinity of the Coulomb barrier. Ahsan and Volya \cite{Ahsan2010} firstly demonstrated it in a one-dimensional model with an exact solution. Later, it was observed in CDCC calculations for deuteron breakup at low energies by Ogata and Yoshida\cite{Ogata2016}. In this section, we not only test the validity of RLLM in the case where the long-range Coulomb coupling is considerable but also investigate the closed channel effect on $^6$Li induced reactions.

Similar to the case in Sec. \ref{appl-6+12}, we calculate the elastic scattering angular distributions for $^6$Li+$^{59}$Co reactions at EL=12.0 and 18.0 MeV and breakup reaction cross section at EL=17.4 MeV. The $\alpha$-$^{59}$Co and $d$-$^{59}$Co optical potentials are taken from Refs. \cite{Avrigeanu2014} and \cite{An2006} respectively. The parameters of $\alpha$ and $d$ optical potentials are obtained at the incident energies 12.0 and 6.0 MeV respectively. The surface imaginary part of $d$-$^{59}$Co optical potential is multiplied by 0.3 to fit the $^6$Li+$^{59}$Co elastic scattering experimental data. The reduction of the imaginary part of $d$ optical potential is due to the strong suppression of $d$ breakup in $^6$Li induced reactions at incident energies around the Coulomb barrier. Watanabe et al. \cite{Watanabe2012} have verified it by comparing the four- and three-body CDCC calculations for $^6$Li elastic scattering and they concluded that $\alpha+d$ model will be good for $^6$Li CDCC calculations if the contribution of $d$ breakup effect is removed from $d$ optical potential. This reduction has been adopted by many researchers \cite{Lei2015,Gomez-Ramos2017,Kumawat2020} in $^6$Li CDCC calculations at relatively low energies. The spin-orbit coupling term of $d$-$^{59}$Co optical potential is ignored as its effect is insignificant.

\begin{table}[tbp]
\caption{Same as Table \ref{table-elas-smat-6+12} but for $^6$Li+$^{59}$Co reaction at EL=18.0 MeV. The $S$-matrices are multiplied by 100. The calculation conditions are: top, $l_{\max}$=4, $N$=25, $R_m$=40 fm; second top, $l_{\max}$=4, $k_{\max}$=2.0 fm$^{-1}$, $R_m$=40 fm; third top, $l_{\max}$=4, $N$=25, $k_{\max}$=2.0 fm$^{-1}$; bottom, $N$=25, $R_m$=40 fm, $k_{\max}$=2.0 fm$^{-1}$. }
\label{table-elas-smat-6+59}
\begin{indented}
\item[]
\begin{tabular}{ccccc}
\br
  $k_{\max}$ (fm$^{-1}$) & 1.0 & 1.6 & 1.8 & 2.0  \\
\mr
 $J^{\pi}$=0$^{-}$  &  5.28-7.67$i$ &   4.67-8.13$i$ &   4.71-8.09$i$ &   4.72-8.03$i$ \\
 $J^{\pi}$=4$^{+}$  &  13.4-5.92$i$ &   13.0-6.45$i$ &   13.1-6.74$i$ &   13.2-6.82$i$ \\
 $J^{\pi}$=11$^{-}$ &  81.7+15.0$i$ &   80.0+15.5$i$ &   80.4+15.1$i$ &   80.5+15.1$i$ \\
\br
 $N$ & 10 & 15 & 20 & 25 \\
\mr
 $J^{\pi}$=0$^{-}$  &  4.84-7.94$i$ &   4.72-8.01$i$ &   4.70-8.02$i$ &   4.72-8.03$i$ \\
 $J^{\pi}$=4$^{+}$  &  12.8-6.89$i$ &   13.2-6.65$i$ &   13.2-6.63$i$ &   13.2-6.82$i$ \\
 $J^{\pi}$=11$^{-}$ &  80.1+15.6$i$ &   80.5+15.6$i$ &   80.5+15.4$i$ &   80.5+15.1$i$ \\
\br
 $R_m$ (fm) & 10 & 20 & 30 & 40 \\
\mr
 $J^{\pi}$=0$^{-}$  &  5.00-6.57$i$ &   4.79-8.01$i$ &   4.73-8.06$i$ &   4.72-8.03$i$ \\
 $J^{\pi}$=4$^{+}$  &  11.6-6.09$i$ &   13.2-6.80$i$ &   13.2-6.83$i$ &   13.2-6.82$i$ \\
 $J^{\pi}$=11$^{-}$ &  88.2+6.14$i$ &   69.3+17.6$i$ &   80.5+15.1$i$ &   80.5+15.1$i$ \\
\br
 $l_{\max}$ & 0 & 2 &  3 & 4\\
\mr
 $J^{\pi}$=0$^{-}$  &  1.50-6.27$i$ &   4.63-7.96$i$ &   4.10-7.40$i$ &   4.72-8.03$i$ \\
 $J^{\pi}$=4$^{+}$  &  7.80-6.61$i$ &   13.2-6.65$i$ &   12.6-6.15$i$ &   13.2-6.82$i$ \\
 $J^{\pi}$=11$^{-}$ &  73.3+18.5$i$ &   80.6+15.2$i$ &   81.1+15.7$i$ &   80.5+15.1$i$ \\
\br
\end{tabular}
\end{indented}
\end{table}

Table \ref{table-elas-smat-6+59} shows some elastic scattering $S$-matrices for $^6$Li+$^{59}$Co reaction at EL=18.0 MeV with different parameters. It is found that larger $R_m$ and $N$ are required to reach convergence compared with $^6$Li+$^{12}$C reaction at EL=168.6 MeV. A noticeable issue is that the elastic scattering $S$-matrix converges at $k_{\max}$=2.0 fm$^{-1}$, which is well above the threshold energy of continuum 14.86 MeV (corresponds to $k_{\max}$=0.97 fm$^{-1}$). The closed channel effect is worthy of consideration in this reaction, and at lower incident energies. On the other hand, although the elastic scattering $S$-matrix is not fully converging when $l_{\max}$=2, the difference between the $S^{J,\pi}_{\beta _0 , \beta _0}$ calculated with $l_{\max}$=2 and 4 is less than 3$\%$. It is expected that the coupling effect of $l$>2 continuum is negligible on elastic scattering for this reaction.

\begin{figure}[tbp]
  \centering
  \includegraphics[width=0.8\textwidth]{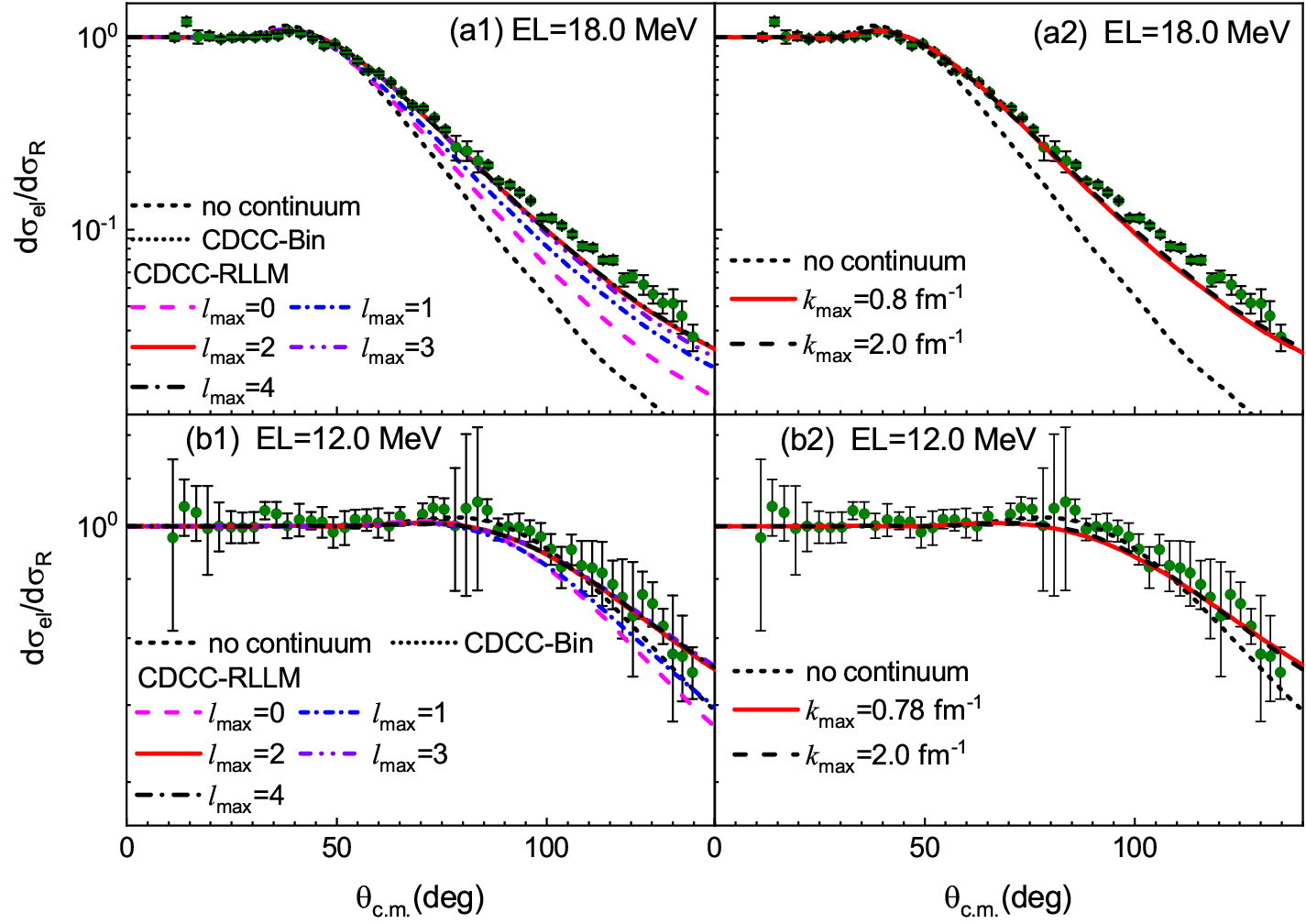}
  \caption{Calculated elastic scattering angular distributions in the Rutherford ratio for $^6$Li+$^{59}$Co reactions at EL=12.0 and 18.0 MeV. The circles are the experimental data taken from Ref \cite{Beck2007}. In subfigures (a1) and (b1), the CDCC-RLLM results with $l_{\max}$=2, 3, 4 and the CDCC-Bin results are almost the same. $k_{\max}$=0.78 fm$^{-1}$ corresponds to the threshold energy of continuum 9.42 MeV for $^6$Li+$^{59}$Co reaction at EL=12.0. See text for details.}
  \label{fig-elastic-6+59}
\end{figure}

Fig. \ref{fig-elastic-6+59} shows the elastic scattering angular distributions of $^6$Li+$^{59}$Co at EL=12.0 and 18.0 MeV calculated with increasing $l_{\max}$ and $k_{\max}$. $N$=25 and $R_m$=40 fm are adopted for each calculation. $k_{\max}$=2.0 fm$^{-1}$ for Fig. \ref{fig-elastic-6+59} (a1) and (b1). $l_{\max}$=2 for Fig. \ref{fig-elastic-6+59} (a2) and (b2). The CDCC results of two reactions are both in good agreement with the experimental data \cite{Beck2007} and they differ from elastic scattering angular distributions calculated without continuum channels, which means that breakup effect is significant whether the incident energy is below or above the Coulomb barrier.

More specifically, the CDCC results at two incident energies both converge with $l_{\max}$=2, which confirms the previous inference. The calculated elastic scattering angular distributions for $^6$Li+$^{59}$Co reactions at EL=18.0 and 12.0 MeV converge at $k_{\max}$=0.8 and 0.78 fm$^{-1}$ respectively. $k_{\max}$=0.78 fm$^{-1}$ corresponds to the threshold energy of continuum 9.42 MeV for $^6$Li+$^{59}$Co reaction at EL=12.0 MeV. The closed channel effect on elastic scattering is invisible whether the incident energy is above or below the Coulomb barrier.

The elastic scattering angular distributions calculated with the bin method are also shown in Fig. \ref{fig-elastic-6+59}. $l_{\max}$=2. $k_{\max}$=2.0 fm$^{-1}$. $R_m$=40 fm. $\Delta k$ is set to be 0.025 fm$^{-1}$ when the momentum $k$ is below 0.7 fm$^{-1}$ and 0.9 fm$^{-1}$ for EL=12.0 and 18.0 MeV respectively. $\Delta k$=0.1 fm$^{-1}$ for higher momentum region. $r_{bin}$=250 fm, which can well ensure the normalization of all $\phi ^{n}_{l,j,I}$ generated by the bin method. The numbers of states for CDCC-Bin are 288 at EL=12.0 MeV and 330 at EL=18.0 MeV, which are both larger than that for CDCC-RLLM in the same CDCC-model space (the number is 118 when $N$=25). Converging elastic scattering angular distributions can be provided by these bin schemes and they are the same as the converging CDCC-RLLM results. It can be concluded that the long-range couplings are handled well by the RLLM.

\begin{figure}[tbp]
  \centering
  \includegraphics[width=0.6\textwidth]{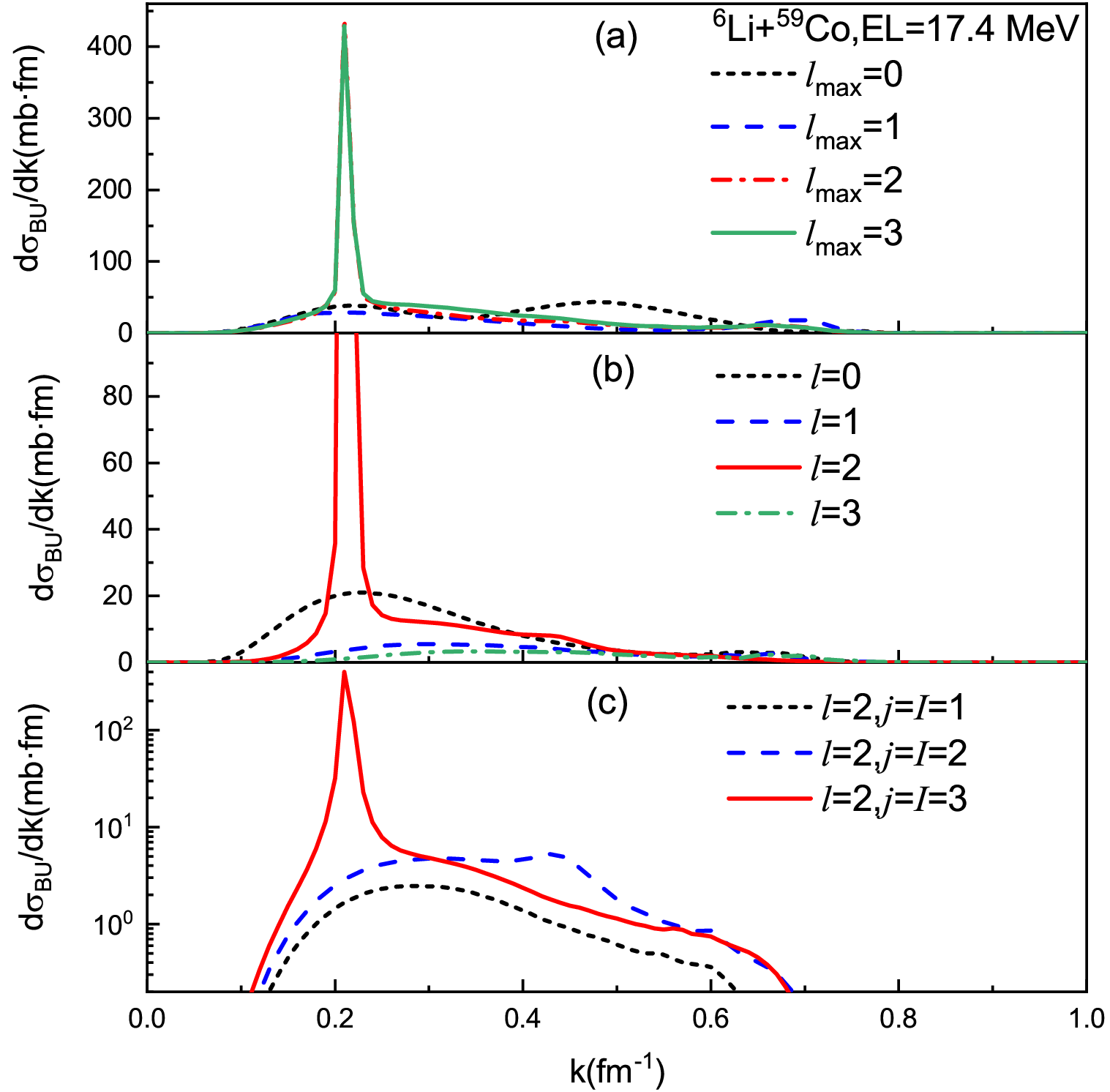}
  \caption{(a)Calculated differential breakup reaction cross sections with different $l_{\max}$ for $^6$Li+$^{59}$Co reaction at EL=17.4 MeV. (b) The components of differential breakup reaction cross section calculated with $l_{\max}$=3. The individual components correspond to different $l$ states. (c) Differential breakup reaction cross sections calculated with $l_{\max}$=3 for $l$=2 states with $j$=$I$=1, 2 and 3 respectively. }
  \label{fig-dbudk-6+59-all}
\end{figure}

For $^6$Li breakup reaction on $^{59}$Co target at EL=17.4 MeV, $N$=60 is required for CDCC-RLLM to obtain converging breakup reaction cross section. With $k_{\max}$=2.0 fm$^{-1}$, the differential breakup reaction cross sections $d\sigma _{BU} /dk$ are calculated with increasing $l_{\max}$ and shown in Fig. \ref{fig-dbudk-6+59-all}. Opposite to the $^6$Li+$^{12}$C reaction at EL=178.0 MeV, increasing $l_{\max}$ from 2 to 3 slightly enhances the differential breakup reaction cross section in 0.25 fm$^{-1}$ $<k<$ 0.45 fm$^{-1}$. Meanwhile, only the peak of 3$^+$ resonance state can be seen clearly in Fig. \ref{fig-dbudk-6+59-all}(c) while the peaks of 2$^+$ and 1$^+$ resonance states disappear. The main parts of breakup reaction cross section come from the continuum states with $l$=0 and 2 while the contributions from the $l$=1 and 3 continuum states can not be neglected. The calculation with $l_{\max}\geq$4 is out of our computational capability but a larger space model would provide a more precise description for this reaction.

\begin{figure}[tbp]
  \centering
  \includegraphics[width=0.8\textwidth]{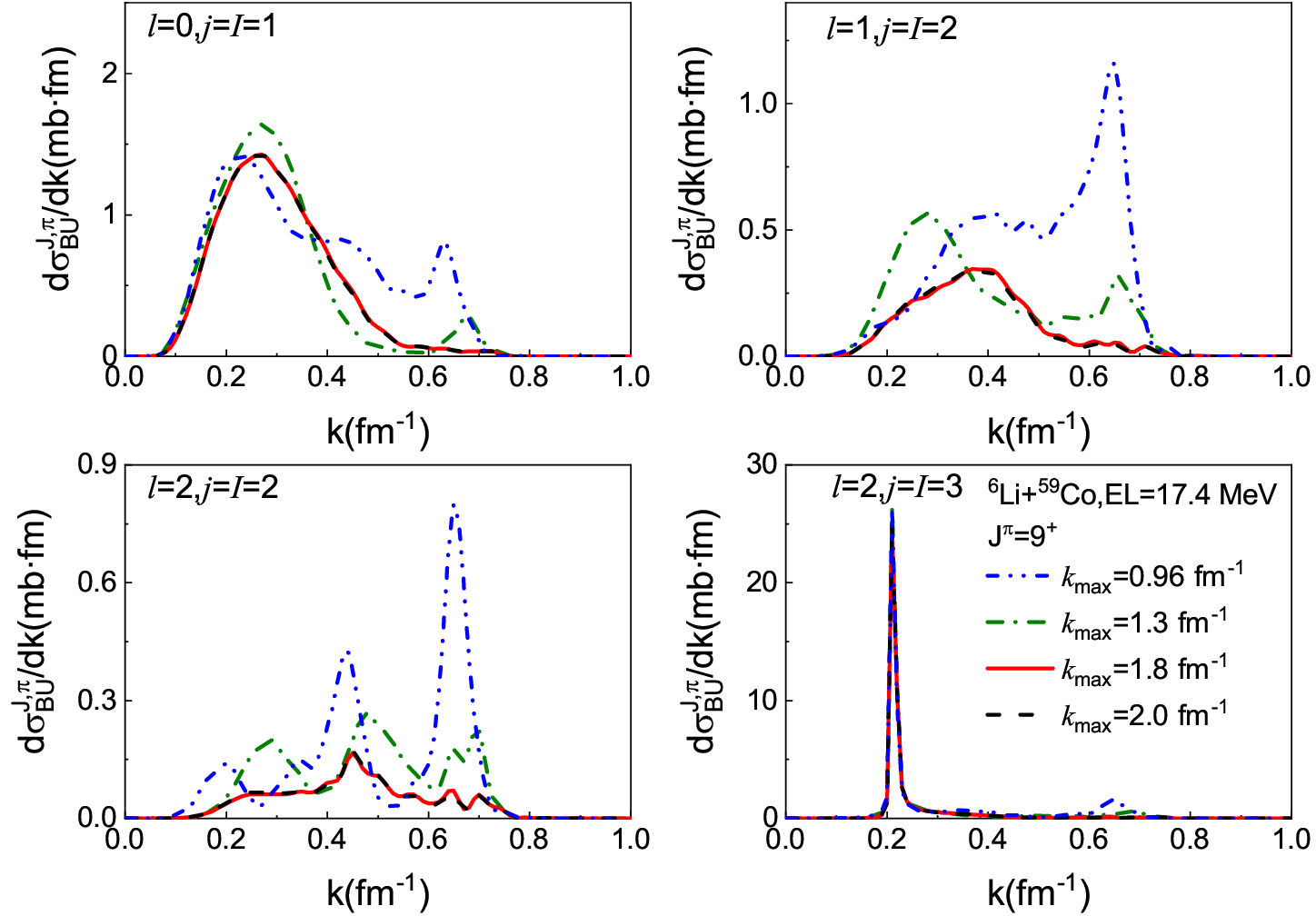}
  \caption{Calculated $d\sigma _{BU}^{J,\pi}/ dk$ by CDCC-RLLM for $^6$Li+$^{59}$Co reaction at EL=17.4 MeV and $J^{\pi}$=9$^+$ with increasing $k_{\max}$. $l_{\max}$=3. $k_{\max}$=0.96 fm$^{-1}$ corresponds to the threshold energy of the continuum 14.31 MeV. See text for details.}
  \label{fig-dbudk-6+59-kmax}
\end{figure}

In order to investigate the closed channel effect on breakup reaction cross sections, we calculate the $d \sigma _{BU}^{J,\pi} / dk$ for some $\alpha -d$ partial waves at $J^{\pi}$=9$^{+}$ with increasing $k_{\max}$ and present them in Fig. \ref{fig-dbudk-6+59-kmax}. $l_{\max}$=3 is adopted. The curves of $k_{\max}$=1.8 fm$^{-1}$ and 2.0 fm$^{-1}$ are almost the same. When $k_{\max}$ is lower than 1.8 fm$^{-1}$, the calculated results are much enhanced, resulting in an overestimation of breakup reaction cross section. The unreasonable peaks at $k \approx $ 0.65 fm$^{-1}$ are removed by increasing $k_{\max}$ from 0.96 to 2.0 fm$^{-1}$. Specifically, the calculated breakup reaction cross sections are 36.84, 27.82, 24.03 and 23.77 mb when $k_{\max}$=0.96, 1.3, 1.8 and 2.0 fm$^{-1}$ respectively. $k_{\max}$=0.96 fm$^{-1}$ corresponds to the threshold energy of the continuum 14.31 MeV for this reaction. Closed channel effect is found to be essential for $^6$Li breakup reaction calculations when the incident energy is around the Coulomb barrier.

\begin{figure}[tbp]
  \centering
  \includegraphics[width=0.6\textwidth]{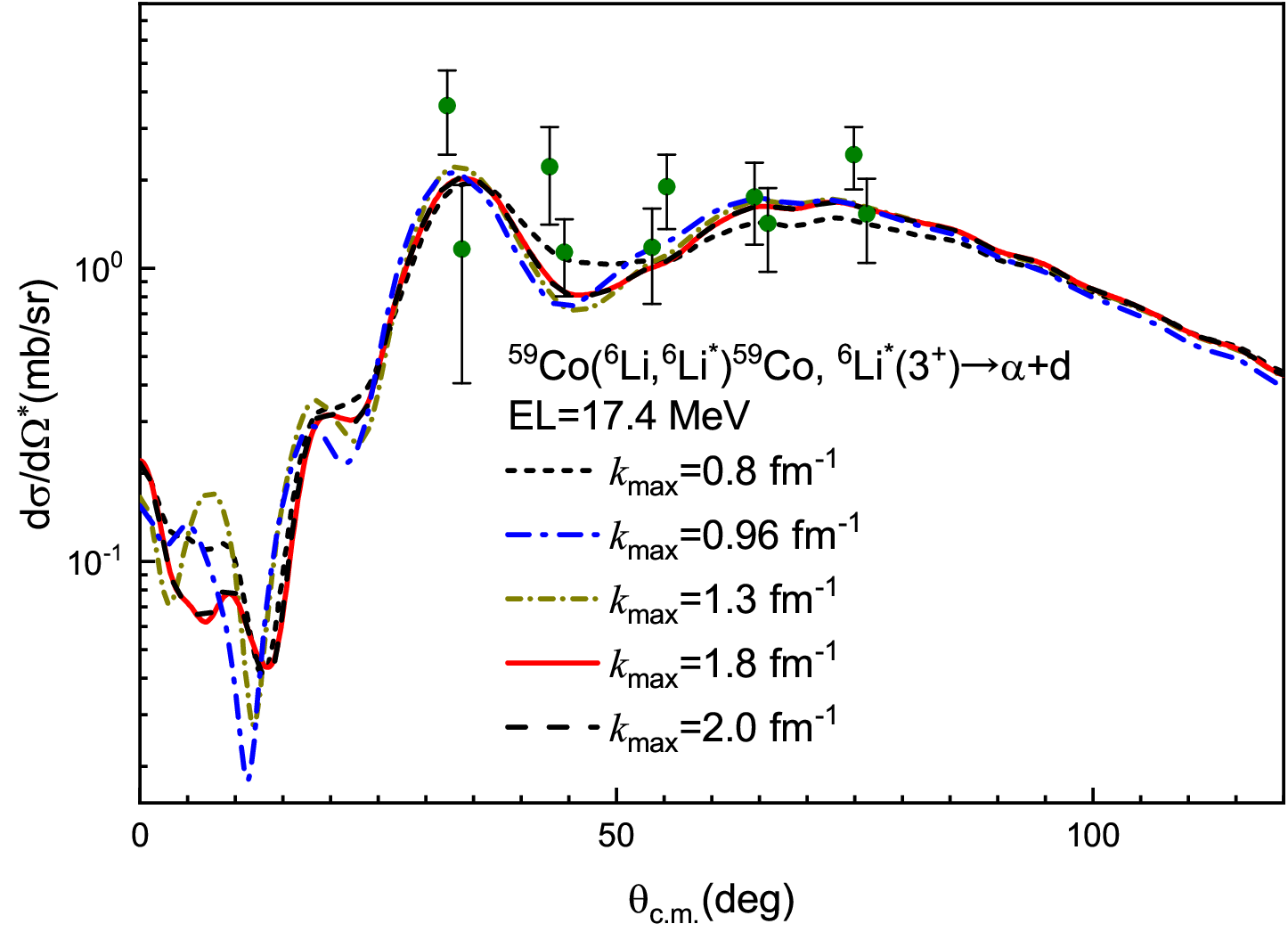}
  \caption{Calculated 3$^+$ resonance state components of exclusive elastic breakup reaction cross sections for $^{59}$Co($^6$Li,$\alpha +d$)$^{59}$Co reaction at EL=17.4 MeV, showing with respect to the scattering angle of the centre of mass of the $\alpha-d$ system. $k_{\max}$=0.96 fm$^{-1}$ corresponds to the threshold energy of the continuum 14.31 MeV. The circles are the experimental data taken from Ref. \cite{Souza2009}.}
  \label{fig-dbudk-6+59-3+}
\end{figure}

Fig. \ref{fig-dbudk-6+59-3+} shows the 3$^+$ resonance state components of exclusive elastic breakup reaction cross sections calculated with different $k_{\max}$. As the errors of experimental data are larger than 20$\%$, all results match the experimental data \cite{Souza2009} well. The results for $k_{\max}$=1.8 fm$^{-1}$ and 2.0 fm$^{-1}$ are almost the same. The calculated angular distribution is reduced visibly between 40-55 degrees and enlarged between 55-90 degrees by increasing $k_{\max}$ from 0.8 fm$^{-1}$ to 0.96 fm$^{-1}$. Further increase of $k_{\max}$ changes the angular distribution visibly in the forward scattering angle region (0-35 degrees) and reduces the calculated result around 60 degrees. The contribution from the 3$^+$ resonance state to breakup reaction is influenced by the closed channels moderately.

\section{Summary and conclusion}\label{sec-conc}

We apply the regularized Lagrange-Laguerre mesh method (RLLM) \cite{Baye2015} to calculate the bound states and discretize the continuum states of weakly bound nuclei for continuum-discretized coupled-channel (CDCC) calculations. With the Gauss quadrature approximation, RLLM is shown to be an efficient and accurate discretization technique for the calculation of elastic scattering and breakup reaction. In the study of $d$+$^{58}$Ni at EL=80.0 MeV, the CDCC-RLLM result is consistent with the previous calculations\cite{Druet2010}. Moreover, for $d$+$^{58}$Ni, $^6$Li+$^{12}$C and $^6$Li+$^{59}$Co at different incident energies, the CDCC-RLLM results are in excellent agreement with the CDCC-Bin results, which verifies the validity of applying Lagrange-mesh method in CDCC. The non-resonance continuum states, resonance continuum states and their effect on scattering can be well handled by RLLM. Particularly, the combination of the Lagrange-mesh method and the improved Numerov algorithm \cite{Yang1980} permits a fast integration of CDCC equations and provides a convenient treatment of closed channels.

Various numerical and physical aspects are discussed for $^6$Li induced reactions. A Woods-Saxon form potential for the $\alpha$-$d$ system is given in the present paper, which can well describe the bound state of $^6$Li and reproduce satisfactorily the resonance energies and resonance widths for the 3$^+$, 2$^+$ and 1$^+$ resonance states of $^6$Li. CDCC calculations based on the combination of the Lagrange-mesh method and the improved Numerov algorithm are performed to the $^6$Li induced reactions. For $^6$Li+$^{12}$C reactions at EL=168.6 and 178.0 MeV, moderate effects are found on elastic scattering angular distribution and breakup reaction cross section when $l$>2 continuum states are included in CDCC calculation. For $^6$Li+$^{59}$Co reactions at incident energies around the Coulomb barrier, the closed channel effect is found to be negligible for elastic scattering angular distribution but that is significant for breakup reaction calculation. A severe overestimation will be made for the breakup reaction cross section if closed channels are excluded from the CDCC equations. So far, CDCC has been widely applied to calculate the total fusion reaction cross section \cite{Beck2007,Camacho2019}, which is estimated by subtracting the breakup reaction cross section from the total reaction cross section. The effect of closed channels on total fusion reaction cross section and corresponding uncertainty will be studied in the future.

Finally, it is noted that RLLM could be extended to describe the bound and continuum states of a three-body system in hyperspherical coordinate \cite{Baye2015}. Therefore, it is worthy of applying RLLM in the study of reactions induced by three-body projectiles, such as $^6$He($\alpha +2n$), $^{11}$Li($^9$Li+2$n$) and $^9$Be(2$\alpha$+$n$). Relative study is in progress.

\ack
This work was supported by the National Natural Science Foundation of China (11705009) and Science Challenge Project (TZ2018005).

\appendix

\section{Prof Yang's improved Numerov algorithm for solving coupled channels equation}\label{appe-Yang}

Yang \cite{Yang1980} improved the coupled channels calculation by using the iteration of the linear relationship between radial wave functions at two neighbouring points. For simplicity, the coupled channels equation is written as
\begin{equation}\label{e-CCeq}
\eqalign{
& \frac{d^2}{dR^2}u_i\left( R \right) =\sum_{j=1}^{N_{ch}}{A_{ij}\left( R \right) u_j\left( R \right)},
\\
& A_{ij}=\left[ \frac{2\mu}{\hbar ^2}\left( \varepsilon _i-E \right) +\frac{L_i\left( L_i+1 \right)}{R^2} \right] \delta _{ij}+\frac{2\mu}{\hbar ^2}V_{ij},
}
\end{equation}
or in matrix form
\begin{equation}\label{e-CCeq-matr}
\frac{d^2}{dR^2}\bm{u}(R)=\bm{A}(R)\bm{u}(R).
\end{equation}
For channel $i$, $\varepsilon _i$ is the channel energy, $L_i$ is the orbital angular momentum. $E$ is the scattering energy. $\mu$ is the reduced mass of the system. $V_{ij}$ is the coupling potential. At asymptotic radius ($R_m$), the wave functions reach their asymptotic behaviours and can be written as
\begin{eqnarray}\label{e-asym}
u_i\left( R \right) &=  I_{L_i}\left( K_iR \right) \delta _{i,\omega}-\sqrt{\frac{v_{\omega}}{v_i}} S_{i,\omega} O_{L_i}\left( K_iR \right) ,E>\varepsilon _i \\
	&= -\sqrt{\frac{v_{\omega}}{v_i}} S_{i,\omega} W_{-\eta _i,L_i+1/2}\left( 2K_ir \right) ,E<\varepsilon _i,
\end{eqnarray}
where $\eta _i$, $v_i$ and $K_i$ is the Sommerfield number, velocity and wave number of channel $i$ respectively ($K_i=\sqrt{2\mu \left| E-E_i \right|}/\hbar $, $v_i=\hbar K_i / \mu$). $\omega$ is the entrance channel. $I_{L_i}(x)$ and $O_{L_i}(x)$ are the incoming and outgoing Coulomb functions \cite{Thompson2010}, and $W_{a,b}(x)$ is the Whittaker function \cite{Olde2010}. $S_{i,\omega}$ is the $S$-matrix.

In general Numerov method, the coupled channels equation is integrated from starting radius ($R_0$) to  $R_m$. $R_0$ is set to be a small radius as that in \emph{Fresco} \cite{Thompson1988} for the stability of integration. The interval $[ R_0,R_m ]$ is discretized by using equally spaced grid, where the step size is $\Delta R=R_{n+1}-R_{n}$. One can obtain the recurrence relation of wave functions at three neighbouring points, that is
\begin{equation}\label{e-Noum-u}
\eqalign{
&\left( 1-\frac{\Delta R^2}{12}\bm{A}_{n-1} \right) \bm{u}_{n-1}-\left( 2+\frac{5}{6}\Delta R^2\bm{A}_n \right) \bm{u}_n \\
&+\left( 1-\frac{\Delta R^2}{12}\bm{A}_{n+1} \right) \bm{u}_{n+1}=0, \\
&\bm{A}_n=\bm{A}\left( R_n \right) ,\bm{u}_n=\bm{u}\left( R_n \right).
}
\end{equation}
It is convenient to rewrite the Eq. (\ref{e-Noum-u}) as the following form
\begin{equation}\label{e-Noum-xi}
  \bm{\xi}_{n-1}+\bm{B}_{n}\bm{\xi}_n+\bm{\xi}_{n+1}=0,
\end{equation}
where
\begin{eqnarray}\label{e-Noum-xi-Bxi}
  \bm{\xi}_n &=\left( 1-\frac{\Delta R^2}{12} \bm{A}_n \right) \bm{u}_n, \\
  \bm{B}_n &=-\frac{2+\frac{5}{6}\Delta R^2 \bm{A}_n}{1-\frac{\Delta R^2}{12} \bm{A}_n}=10-\frac{12}{1-\frac{\Delta R^2}{12} \bm{A}_n}.
\end{eqnarray}

The seven-point starting formula can be adopted in Yang's method, that is
\begin{equation}\label{e-Yang-starting}
\eqalign{
\bm{u}_0-\left( 1+\frac{67}{48}\Delta R^2 \bm{A}_1 \right) \bm{u}_1+\frac{\Delta R^2}{6} \bm{A}_2u_2-\frac{61}{24}\Delta R^2 \bm{A}_3 \bm{u}_3+\frac{\Delta R^2}{6} \bm{A}_4 \bm{u}_4
\\
-\left( 1+\frac{67}{48}\Delta R^2 \bm{A}_5 \right) \bm{u}_5+\bm{u}_6=0.
}
\end{equation}
At the starting point, $\bm{u}_0$=$\bm{u}(R_0)$=0. Using Eq. (\ref{e-Noum-xi-Bxi}), Eq. (\ref{e-Yang-starting}) can be written as
\begin{equation}\label{e-Yang-starting-xi}
 \bm{P}_1 \bm{\xi} _1-\bm{P}_2 \bm{\xi} _2+\bm{P}_3 \bm{\xi} _3 -\bm{P}_4 \bm{\xi} _4+\bm{P}_5 \bm{\xi} _5-\bm{P}_6 \bm{\xi} _6=0.
\end{equation}
where $\bm{P}_n=a_n+b_n \bm{B}_n$, $n$=1,2,...,6. The coefficients $a_n$ and $b_n$ are given in Table \ref{table-an-bn}.

\begin{table}[tbp]
\centering
\caption{The coefficients $a_n$ and $b_n$ of $\bm{P}_n$ at $n$=1,2,...,6. }
\label{table-an-bn}
\begin{indented}
\item[]
\begin{tabular}{ccccccc}
\br
 $n$  & 1 & 2 & 3 & 4 & 5 & 6 \\
\mr
$a_n$ & 47/24 & 1/3 & 61/12 & 1/3 & 47/24 & -5/6 \\
$b_n$ & 71/48 & 1/6 & 61/24 & 1/6 & 71/48 & 1/12 \\
\br
\end{tabular}
\end{indented}
\end{table}

Starting from Eq. (\ref{e-Yang-starting-xi}) and using Eq. (\ref{e-Noum-xi}) to eliminate the $\bm{\xi}_1$, $\bm{\xi}_2$, $\bm{\xi}_3$, ..., and so on, one can get the recurrence relation for $\bm{Q}_n$,
\begin{eqnarray}\label{e-recc-Q}
\bm{Q}_1 &= \bm{P}_1
\\
\bm{Q}_2 &=\bm{Q}_1 \bm{B}_1+\bm{P}_2
\\
\bm{Q}_n &=\bm{Q}_{n-1} \bm{B}_n-\bm{Q}_{n-2}+\bm{P}_n,n=3,4,5,6
\\
\bm{Q}_n &=\bm{Q}_{n-1} \bm{B}_n-\bm{Q}_{n-2}, n \geq 7,
\end{eqnarray}
and the relation for the two neighbouring functions
\begin{equation}\label{e-rela-xi}
  \bm{Q}_n \bm{\xi} _n + \bm{Q}_{n-1} \bm{\xi} _{n+1}=0
\end{equation}

At asymptotic radius $R_m=R_0+m\Delta R$, the coupling potential reduces to
\begin{equation}\label{e-asym-Vcc}
  V_{ij}\longrightarrow \frac{Z_P Z_T e^2}{R} \delta_{ij}
\end{equation}
where $Z_P$ and $Z_T$ are the charge numbers of projectile and target respectively. Therefore the off-diagonal part of $\bm{A}$ can be neglected and the $S$-matrix can be obtained by solving linear equation
\begin{equation}\label{e-Smat-eq}
  \bm{C}\bm{S}_{\omega}=\bm{D},
\end{equation}
where
\begin{eqnarray}\label{e-Smat-eq-CDS}
& (\bm{S}_{\omega})_i=S_{i,\omega}, \\
& (\bm{C})_{ij}=\sqrt{\frac{v_{\omega}}{v_j}}\left[ (\bm{X}^{(1)})_{ij}(\bm{O}^{(1)})_{j} +(\bm{X}^{(2)})_{ij}(\bm{O}^{(2)})_{j} \right] \\
& (\bm{D})_i=(\bm{X}^{(1)})_{i\omega} (\bm{I}^{(1)})_{\omega}+(\bm{X}^{(2)})_{i\omega} (\bm{I}^{(2)})_{\omega}, \\
& \bm{X}^{(1)}=\bm{Q}_{m-1}\left( 1-\frac{h^2}{12} \bm{A}_{m-1} \right), \\
& \bm{X}^{(2)}=\bm{Q}_{m-2}\left( 1-\frac{h^2}{12} \bm{A}_{m} \right), \\
& (\bm{I}^{(1)})_i=I_{L_i}(K_iR_{m-1}) \delta_{i,\omega}, \\
& (\bm{I}^{(2)})_i=I_{L_i}(K_iR_{m}) \delta_{i,\omega}.
\end{eqnarray}
For open channel ($E>\varepsilon _i$),
\begin{eqnarray}\label{e-O-open}
& (\bm{O}^{(1)})_i=O_{L_i}(K_iR_{m-1}) , \\
& (\bm{O}^{(2)})_i=O_{L_i}(K_iR_{m}) ,
\end{eqnarray}
and for closed channel ($E<\varepsilon _i$),
\begin{eqnarray}\label{e-O-close}
& (\bm{O}^{(1)})_i=W_{-\eta _i,L_i+1/2}(2K_iR_{m-1}), \\
& (\bm{O}^{(2)})_i=W_{-\eta _i,L_i+1/2}(2K_iR_{m}).
\end{eqnarray}

\section{A stabilization method for Prof Yang's algorithm}\label{appe-Stable}

For the general Numerov method, Tolsma and Veltkamp \cite{Tolsma1986} have pointed out that an exponentially growing part and an exponentially decreasing part will be produced as $R$ increases through a classically forbidden region, where the local kinetic energy is negative. The former one reduces the linear dependence of the coupled channels solutions and leads to the failure of solving coupled channels equation.

In Yang's algorithm, $\bm{Q}_n$ determines the linear relationship between radial wave functions as shown in Eq. (\ref{e-rela-xi}). As the equation is integrated through a classically forbidden region, the exponentially growing part will be the main part of $\bm{\xi}$ and the exponentially decreasing part will be little. Therefore, $Q_n$ can not be applied for the latter one, which is the required function.

Baylis et al. \cite{Baylis1982} and \emph{Fresco} \cite{Thompson1988} provide an re-orthogonalization procedure for general Numerov method to improve the linear dependence of the coupled channels solutions, which is based on $\bm{\xi}$. Similarly, we give a stabilization method for Prof Yang's algorithm, which is based on $\bm{Q}_n$, however. The transposed matrix of $\bm{Q}_n$ is decomposed by QR factorization as
\begin{equation}\label{e-appe-QR}
  ( \bm{Q}_n ) ^T=\bm{Q}_{n}^{QR}\bm{R}_{n},
\end{equation}
where $\bm{Q}_{n}^{QR}$ is an orthogonal matrix and $\bm{R}_{n}$ is an upper triangular matrix. Then we use $(\bm{R}_{n}^{-1})^T\bm{Q}_{n-1}$ and $(\bm{Q}_{n}^{QR})^T$ to replace $\bm{Q}_{n-1}$ and $\bm{Q}_{n}$ and continue the recurrence of $\bm{Q}$. This procedure can be done periodically to ensure the stability of the solution. In Sec. \ref{appl-6+59}, this procedure is necessary when closed channels are included in CDCC equations.

\section*{References}
\bibliography{mythesis}
\end{document}